\documentclass[fleqn,usenatbib,useAMS,useAMS,usenatbib]{mnras}

\usepackage{hyperref}
\usepackage{xcolor}
\usepackage{color}
\usepackage{float}
\usepackage{subfigure}
\usepackage{tabularx}
\usepackage{soul}
\usepackage{comment}

\usepackage{graphicx}	
\usepackage{amsmath}	
\usepackage{amssymb}	
\usepackage{multicol}        
\usepackage{amsfonts}
\usepackage{pdflscape}	

\usepackage[T1]{fontenc}
\usepackage{ae,aecompl}

\usepackage{newtxtext,newtxmath}
\usepackage{bm}	
\usepackage{graphicx}
\usepackage{dcolumn}
\usepackage{amsmath}
\usepackage{amsfonts}
\usepackage{amssymb}
\usepackage{color}
\usepackage{float}
\usepackage{subfigure}
\usepackage{tabularx}
\usepackage{soul}
\usepackage{mathtools}
\usepackage{cuted}

\newcolumntype{L}[1]{>{\raggedright\arraybackslash}p{#1}}
\newcolumntype{C}[1]{>{\centering\arraybackslash}p{#1}}
\newcolumntype{R}[1]{>{\raggedleft\arraybackslash}p{#1}}

\definecolor{v}{rgb}{0.6, 0.2, 0.8} 

\providecommand{\U}[1]{\protect\rule{.1in}{.1in}}

\usepackage{tikz,xcolor,hyperref}

\definecolor{lime}{HTML}{A6CE39}
\DeclareRobustCommand{\orcidicon}{%
	\begin{tikzpicture}
	\draw[lime, fill=lime] (0,0) 
	circle [radius=0.16] 
	node[white] {{\fontfamily{qag}\selectfont \tiny ID}};
	\draw[white, fill=white] (-0.0625,0.095) 
	circle [radius=0.007];
	\end{tikzpicture}
	\hspace{-2mm}
}

\foreach \x in {A, ..., Z}{%
	\expandafter\xdef\csname orcid\x\endcsname{\noexpand\href{https://orcid.org/\csname orcidauthor\x\endcsname}{\noexpand\orcidicon}}
}

\usepackage{hyperref}
\hypersetup{
    colorlinks=true,
    linkcolor=red,
    citecolor=blue,
    filecolor=magenta,      
    urlcolor=cyan,
    pdftitle={Cosmology under the fractional calculus approach},
    pdfpagemode=FullScreen} 

\begin{document}

\title{Cosmology under the fractional calculus approach}

\author[Garc\'ia-Aspeitia, Fernandez-Anaya, Hern\'andez-Almada,Leon, Maga\~na]{Miguel A. Garc\'ia-Aspeitia\orcidA{}$^1$ \thanks{E-mail: angel.garcia@ibero.mx}, Guillermo Fernandez-Anaya\orcidB{}$^1$ \thanks{E-mail: guillermo.fernandez@ibero.mx},   A.  Hern\'andez-Almada\orcidC{}$^2$ \thanks{E-mail: ahalmada@uaq.mx}, \newauthor Genly Leon\orcidD{}$^{3,4}$ \thanks{E-mail: genly.leon@ucn.cl},  Juan Maga\~na\orcidE{}$^{5}$
\thanks{E-mail: juan.magana@ucentral.cl} \\
$^1$ Depto. de F\'isica y Matem\'aticas, Universidad Iberoamericana Ciudad de M\'exico, Prolongaci\'on Paseo \\ de la Reforma 880, M\'exico D. F. 01219, M\'exico
\\$^2$ Facultad de Ingenier\'ia, Universidad Aut\'onoma de
Quer\'etaro, Centro Universitario Cerro de las Campanas, 76010, \\
Santiago de Quer\'etaro, M\'exico
\\$^3$ Departamento  de  Matem\'aticas,  Universidad Cat\'olica del Norte, Avda.   Angamos  0610,  Casilla  1280  Antofagasta,  Chile\\
$^4$  Institute of System Science, Durban University of Technology, PO Box 1334,
Durban, 4000, South Africa\\
$^5$ Escuela de Ingenier\'ia, Universidad Central de Chile, Avenida Francisco de Aguirre 0405, 171-0164 La Serena, Coquimbo, Chile
}

\maketitle
\begin{abstract}
Fractional cosmology modifies the standard derivative to Caputo's fractional derivative of order $\mu$, generating changes in General Relativity. Friedmann equations are modified, and the evolution of the species densities depends on $\mu$ and the age of the Universe $t_U$. We estimate stringent constraints on $\mu$ using cosmic chronometers, Type Ia supernovae, and joint analysis. We obtain $\mu=2.839^{+0.117}_{-0.193}$ within the $1\sigma$ confidence level providing a non-standard cosmic acceleration at late times; consequently, the Universe would be older than the standard estimations. Additionally, we present a stability analysis for different $\mu$ values. This analysis identifies a late-time attractor corresponding to a power-law decelerated solution for $\mu < 2$. Moreover, a non-relativistic critical point exists for $\mu > 1$ and a sink for $\mu > 2$. This solution is a decelerated power-law if $1 < \mu < 2$ and an accelerated power-law solution if $\mu > 2$, consistent with the mean values obtained from the observational analysis. 
Therefore, for both flat FLRW and  Bianchi I metrics, the modified Friedmann equations provide a late cosmic acceleration under this paradigm without introducing a dark energy component. This approach could be a new path to tackling unsolved cosmological problems.
\end{abstract}


\date{Accepted 2022 October 15. Received 2022 October 12; in original form 2022 July 19} 
\pubyear{2022}
\label{firstpage}
\pagerange{\pageref{firstpage}--\pageref{lastpage}}
\section{Introduction}
Modern background cosmology is based on diverse hypotheses, such as the species of fluids. Those species are baryonic matter, photons, neutrinos, and the elusive and mysterious dark matter (DM) and dark energy (DE). In particular, the D.E. component in standard cosmology the well-known $\Lambda$CDM model) is considered a cosmological constant ($\Lambda$). This mentioned model has several achievements and helps us to describe the late time acceleration observed by Supernovas of the Ia type (SnIa) \citep{Riess:1998} and confirmed by the Cosmic Microwave Background radiation (CMB) \citep{Planck:2018}. On the other hand, $\Lambda$CDM describes the structure formation, with an excellent concordance with observations, assuming the presence of cold D.M.
Despite these achievements, there are several cracks in their physical and mathematical structure, like the inability to quantify the quantum vacuum fluctuations when we interpret the $\Lambda$ in this way \citep{Zeldovich, Weinberg}. Additionally, the origin of the late time acceleration of the Universe remains unknown \citep{Carroll:2000}. On the other hand, the Hubble constant value measured with local observations (see SH0ES \citet{Riess:2019cxk}) is in tension with that estimated from early observations (see Planck \citet{Planck:2018}). A possible alternative to solve this tension is to consider extensions beyond $\Lambda$CDM (see \citet{DiValentino:2021izs} for a compilation). However, incomprehension between the SnIa absolute magnitude and the Cepheid-based distance ladder instead of an exotic late-time physics could be the reason for the $H_0$ tension \citep{Efstathiou:2021ocp}.

The community searches for extensions to the $\Lambda$CDM model to resolve some of the mentioned problems. The approaches to face them are divided into two main branches: i) assume a DE fluid with the capability to accelerate the Universe or ii) modify General Relativity (GR) to obtain the cosmic acceleration without DE \citep{Motta:2021hvl}. This paper will focus on the second point under the formalism known as {\it fractional calculus}, which consists of a generalization of the classical integer order calculus, whose derivatives and integrals are of (real or complex) arbitrary order. These fractional operators are not local. In many cases can model real-world phenomena in a better fashion than those obtained by classical calculation. For example, it is coined fractional dynamics as a field of study in physics and mechanics investigating the behavior of objects and systems that are characterized by power-law nonlocality, power-law long-term memory or fractal properties by us and differentiation of non-integer orders, i.e., by methods in the fractional calculus (see the review \citet{Tarasov2013}). Fractional calculus is a field with multiple applications and a great deal of research activity. Fractional quantum mechanics is employed as a tool within quantum field theory and gravity for fractional spacetime \citep{Calcagni:2010bj,Calcagni:2009kc} and the fractional quantum field theory at positive temperature \citep{Lim:2006hp,LimEab+2019+237+256} and other applications of quantum cosmology \citep{VargasMoniz:2020hve,Moniz:2020emn,Rasouli:2021lgy,Jalalzadeh:2021gtq}. Recently, the community explores the fractional calculus to tackle problems associated in cosmology \citep{Shchigolev:2010vh,Shchigolev:2012rp,Shchigolev:2013jq,Calcagni:2013yqa,Shchigolev:2015rei,Calcagni:2016azd,Shchigolev:2021lbm,Jalalzadeh:2022uhl,Calcagni:2020ads,Calcagni:2021ipd,Calcagni:2021aap},  stochastic GW background \citep{Calcagni:2020tvw}, luminosity distance \citep{Calcagni:2019ngc}, inflation and CMB spectrum \citep{Calcagni:2017via,Calcagni:2016ofu}, Fractional Action Cosmology \citep{El-Nabulsi:2012wpc,El-Nabulsi:2015szp,Jamil:2011uj}, 
 fractional geodesic equation, complex general relativity, and discrete gravity \cite{El-Nabulsi:2013mma}, minimal couplings \citep{El-Nabulsi:2013mwa}, phantom \citep{Rami:2015kha}, Ornstein-Uhlenbeck-like fractional differential equation in cosmology \citep{El-Nabulsi:2016dsj}, a variable Order Parameter \citep{El-Nabulsi:2017vmp}, wormholes in fractional action cosmology \citep{El-Nabulsi:2017jss}. New metrics were considered \citep{El-Nabulsi:2017rdu}, as well as some   dark energy models in emergent, logamediate, and intermediate scenarios of the universe \citep{Debnath2012,Debnath2013}. For instance, \citet{Shchigolev:2015rei,Shchigolev:2021lbm} found $\alpha=0.926$ (where $\alpha$ is the order of the Riemann-Liouville fractional integral).  In \citet{Shchigolev:2010vh, Shchigolev:2012rp, Shchigolev:2013jq} were obtained several exact solutions for cosmological models, which differs significantly from the standard model due to the fractal nature of spacetime \citep{Calcagni:2010bj,Calcagni:2009kc}. \citet{Jalalzadeh:2022uhl} explore the interval $1<\alpha <2$ using Riesz's fractional derivative (that is not related to the index of Riemann-Liouville fractional integral) to obtain the non-boundary and tunneling wave functions for a closed de Sitter geometry. Another example, \citet{Barrientos:2020kfp}, studies the Universe dynamics without DM and DE components by modifying the mathematical structure of Friedmann equations with fractional calculus. Another approach is calculating the value of the $\Lambda$ (due to the well-known ultraviolet divergence in the standard quantum field theory), which needs to restructure the theory using the fractional calculus \citep{Calcagni_2021}. In, \citet{Giusti:2020rul, Torres:2020xkw} explore Modified Newtonian Dynamics Theories (MOND) and quantum cosmology in this fractional approach \citep{Barrientos:2020kfp}. Finally, notice that there are several definitions of fractional derivatives and fractional integrals, such as those of Riemann-Liouville, Caputo, Riesz, Hadamard, Marchand, and Griinwald-Letnikov, among other more recent ones (see \citet{book1:2006}, and \citet{book2:1999}  and references therein). Even though these operators are already well studied, some of the usual features related to function differentiation fails, such as Leibniz's rule, the chain rule, and the semi-group property \citep{book2:1999,book1:2006}. 

Based on the fractional calculus formalism, a modified Friedmann equation is confronted with data at the background cosmology. In particular, we use Cosmic Chronometers, Type Ia Supernovae observations, and a joint analysis to constrain the fractional parameter. We will show that the term containing the fractional parameter act as a $\Lambda$, unveiling that nature can be fractional, and consequently, non-fractional GR is only an approach to the actual mathematical structure of nature. Additionally, we present a dynamical system and stability analysis to explore the phase space for different values of the fractional parameter. Finally, we introduce relevant variables for the model and solve the Friedman restriction locally around the equilibrium points, obtaining a reduced phase plane. Finally, we classify these equilibrium points and provide a range on the fractional parameter to obtain a late-term physical accelerated power-law solution for the scale factor. 

The paper is organized as follows: In Sec. \ref{MF}, we present the mathematical formalism of fractional calculus. In Sec. \ref{sec:cosmology}, we show the cosmology based on this theory and how the fractional term could act as a $\Lambda$. In Sec. \ref{sec:constraints}, we show the data and methodology we will use to constrain the theory's free parameters. In Sec. \ref{Results}, we present the results obtained through the different observations and the joint analysis. In Sec. \ref{DS}, we present a dynamical system and stability analysis of the fractional model. In  Sec. \ref{Bianchi} we examine the Bianchi I cosmology, presenting a phase space analysis. Finally, in Sec. \ref{SD}, we give a summary and final discussions. Finally, we will use units where $\hbar=c=k_{B}=1$ unless we mention otherwise.

\section{Mathematical Formalism for Fractional Calculus} \label{MF}

Currently, several definitions of fractional derivatives \citep{Uch:2013}, like the Riemann-Liouville derivative (R.L.D.), and the Caputo derivative (CD), among others, are used. These derivatives are defined by Cauchy's formula for the integral multiple of integer order $\alpha>0$, in the form
\begin{equation}
{ }_{c} I_{t}^{\alpha} f(t)=\Gamma(\mu)^{-1} \int_{c}^{t} f(\tau)(t-\tau)^{\alpha-1} d \tau \label{EQ1}.
\end{equation}
In this case, the R.L.D. with $\alpha \geq 0$ for $f(t)$ is defined by 
\begin{eqnarray}
D_{t}^{\alpha} f(t) &\equiv & \frac{d^n}{dt^n}  \left({ }_{c}I_{t}^{n-\alpha} f(t)\right)\nonumber\\ &=& \Gamma(n-\alpha)^{-1} \frac{d^{n}}{d t^{n}} \int_{c}^{t} \frac{f(\tau)}{(t-\tau)^{\alpha-n+1}} d \tau, \label{EQ2}
\end{eqnarray}
where $n=[\alpha]+1$ being $\alpha \in (n-1,n)$. Notice that the main parameter of fractional calculus is given by $\alpha$, recovering standard calculus when $\alpha\to1$.
 The Caputo left derivative is defined as
 \begin{align}
 & { }^{C} D_{t}^{\mu} f(t) \equiv { }_{c} I_{t}^{n-\mu} D_{t}^{n} f(t) =    \Gamma(n-\mu)^{-1} \int_{c}^{t} \frac{\frac{d^{n}}{d \tau^{n}} f(\tau)}{(t-\tau)^{\mu-n+1}} d \tau
\end{align}
where $n=\left\{\begin{array}{cc}
     [\mu] +1& \mu \notin \mathbb{N}\\
     \mu &  \mu \in \mathbb{N}
\end{array} \right.$. To differentiate the Caputo's fractional constant with the other fractional theories, we denote the fractional constant with the Greek letter $\mu$ instead of $\alpha$. 

In fractional calculus, we now have the following relation (see \citep{Uch:2013}) for the case of more than one derivatives 
\begin{equation}
D_{t}^{\mu} \left[D_{t}^{\beta} f(t)\right]=D_{t}^{\mu+\beta} f(t)-\sum_{j=1}^{n} D_{t}^{\beta-j} f(c+) \frac{(t-c)^{-\mu-j}}{\Gamma(1-\mu-j)},
\end{equation}
or in other words $D_{t}^{\mu} D_{t}^{\beta} f(t) \neq D_{t}^{\mu+\beta} f(t)$, if only not all derivatives $D_{t}^{\beta-j} f(c+)$  are equal to zero at $c$. 
Additionally, the fractional derivative of the Leibniz rule \citep{Uch:2013} reads as 
\begin{equation}
D_{t}^{\mu}[f(t) g(t)]=\sum_{k=0}^{\infty} \frac{\Gamma(\mu+1)}{k ! \Gamma(\mu-k+1)} D_{t}^{\mu-k} f(t) D_{t}^{k} g(t),
\end{equation}
having the usual when $\mu=n\in \mathbb{N}$.

Finally, we need to mention that in \citep{Shchigolev:2010vh} and \citep{ Roberts:2009ix}, the Riemann curvature tensor and the Einstein tensor are defined as usual, but now with dependence on the $\mu$ fractional parameter. In this vein, it is possible to write down the fractional analogous for the Einstein field equation through the expression
\begin{equation}
G_{\alpha \beta}(\mu)=8 \pi G T_{\alpha \beta}(\mu),
\end{equation}
where $G_{\alpha \beta}(\mu)$ is the Einstein tensor in fractional calculus, and $G$ is the Newton gravitational constant.

In this case, modifications to several astrophysical and cosmological events can be studied. For example, a fractional theory of gravitation for fractional spacetime is developed in \citep{Vacaru:2010fb,Vacaru:2010wj}). Non-holonomic deformations to cosmology lead to new classes of cosmological models studied in \citep{Vacaru:2010wn, Shchigolev:2021lbm}. 

\section{Background Fractional Cosmology} \label{sec:cosmology}
To construct the Lagrangian dynamics, one uses the Fractional Action-Like Variational Approach  developed by \citet{El-Nabulsi:2005oyu, El-Nabulsi:2007lla,El-Nabulsi:2007wgc}, \citet{El-Nabulsi:2008oqn}; and one of the possible versions  by \citet{Roberts:2009ix}.

\subsection{Fractional Action-Like Variational Approach}

The background cosmology is based on the Friedmann-Lema\^{i}tre-Robertson-Walker (FLRW) metric which is written in the form $ds^2=-N^2(t) dt^2+ a^2(t)(dr^2+r^2d\Omega^2)$ in where it is considered a flat Universe ($k=0$) based on Planck observations \citep{Planck:2018}, $a(t)$ is the scale factor and $d\Omega^2\equiv d\theta^2+\sin^2\theta d\varphi^2$ is the solid angle. 
The fractional effective action can be written as
\begin{small}
\begin{align}
 S_{\text{eff}} 
   & = \frac{1}{\Gamma(\mu)}\int_0^t \Bigg[\frac{3}{8\pi G}\Bigg(\frac{a^2(\tau)\ddot{a}(\tau)}{N^2(\tau)}+\frac{a(\tau)\dot{a}^2(\tau)}{N^2(\tau)}-\frac{a^2(\tau)\dot{a}(\tau)\dot{N}(\tau)}{N^3(\tau)}\Bigg) \nonumber\\
    & \;\;\;\;\;\;\;\;\;\;\;\;\;\;\;\;\;\;\;\;\;\;\;\;\;\; +a^3(\tau)\mathcal{L}_{\text{m}}\Bigg](t-\tau)^{\mu-1} N(\tau) d\tau, \label{LFC}
\end{align}
\end{small}
\noindent
where $\Gamma(\mu)$ is the  Gamma function, $\mathcal{L}_{\text{m}}$ is the matter Lagrangian, $\mu$ is the fractional constant parameter, $t$ and $\tau$ are the physical and intrinsic time respectively and where the $\Lambda$ is not considered (see \citet{Shchigolev:2010vh}). Varying the action \eqref{LFC} for $q_i\in \{N, a \}$, we obtain  the Euler-Poisson (EP) equations
from which are deduced the field equations with a gauge $N=1$.

The minimization of matter Lagrangian for the energy-momentum tensor for perfect fluid given by $T_{\alpha \beta}=pg_{\alpha \beta}+(\rho+p)u_{\alpha}u_{\beta}$, where $p$, $\rho$ and $u_{\alpha}$ are pressure, energy density and four-velocity respectively, related through the equation of state (EoS) $w$ as $p=w\rho$, leads to modified Einstein field equations with a perfect fluid source. 

Incorporating the several matter sources, the minimization of the fractional action \eqref{LFC} generates 
the following Raychaudhuri equation (with $N(t)=1$)
 \begin{equation}
   \dot{H}+\frac{(\mu -1) H}{2 t}+\frac{(\mu -2) (\mu -1)}{2 t^2}=- 4 \pi  G \sum_i (p_i+\rho_i), \label{IntegrableH}
\end{equation}
and the Friedmann equation, written in the form
\begin{equation}
    H^2+\frac{(1-\mu)}{t}H=\frac{8\pi G}{3}\sum_i\rho_i, \label{FriedmannFrac}
\end{equation}
where the sum is over all the species, in this case, matter and radiation. 
To designates the independent time variables we use the rule $t-\tau =T \mapsto t$ \citep{Shchigolev:2010vh}, where the dots denotes these derivatives. Additionally, the Hubble parameter is defined as $H\equiv\dot{a}/a$.
Notice that we are considering that does not exist a $\Lambda$ and thus, the additional $(1-\mu)H t^{-1}$-term of the previous equation should generate the late accelerated expansion.

Moreover, the continuity equation is 
\begin{equation}
    \sum_i\left[\dot{\rho}_i+3\left(H+\frac{1-\mu}{3t}\right)(\rho_i+p_i)\right]=0.\label{CE}
\end{equation}
Notice that when $\mu=1$ in Eq. \eqref{FriedmannFrac} and \eqref{CE}, the standard cosmology is recovered without $\Lambda$. 

Using the equation of state $p_i=w_i\rho_i$, where $w_i \neq -1$ are constants, thus we have 
 \begin{align}
       & \sum_i (1+w_i)\rho_i \left[\frac{\dot{\rho}_i}{ (1+w_i) \rho_i} + 3\frac{\dot{a}}{a}+ \frac{1-\mu}{t} \right]\nonumber\\
       &   =  \sum_i (1+w_i)\rho_i  \frac{\mathrm{d}}{\mathrm{d}t}\left[ \ln \left( {\rho_i}^{ 1/(1+w_i)} a^3 t^{1-\mu}\right)\right].
 \end{align}
Assuming separated conservation equations for each species considered in the cosmology
we have the following equation in differential form,
\begin{align}
\mathrm{d}\left[ \ln \left( {\rho_i}^{ 1/(1+w_i)} a^3 t^{1-\mu}\right)\right]   =0. \label{CE2}
\end{align}
Setting $a(t_U)=1$, where $t_U$ is the age of the Universe, and denoting by $\rho_{0i}$ the current value of energy density of the $i$-th species, and integrating Eq. \eqref{CE2}, we have for each of the species the energy densities
\begin{equation}
   \rho_{i}(t)= \rho_{0i} a(t)^{-3 (1+w_i)} \left(t/t_U\right)^{(\mu-1)(1+w_i)}. \label{EQ16}
\end{equation}
Then, substituting \eqref{EQ16} in \eqref{FriedmannFrac}, we obtain 
\begin{align}
   & H^2+\frac{(1-\mu)}{t}H =\frac{8\pi G}{3}\sum_i \rho_{0i} a^{-3 (1+w_i)}\left(t/t_U\right)^{(\mu-1)(1+w_i)}. \label{EH}
\end{align}
To compare with the standard model we impose that the universe components are matter ($\rho_{1}=\rho_{\text{m}}, w_{\text{m}}=0$) and radiation  ($\rho_{2}=\rho_{\text{r}}, w_{\text{r}}=1/3$), which in our modified scenario  evolve  as 
\begin{align}
& \rho_{\text{m}}=\rho_{0\text{m}}a^{-3} \left(t/t_U\right)^{\mu-1}, \; 
 \rho_{\text{r}}=\rho_{0\text{r}}a^{-4}  \left(t/t_U\right)^{\frac{4}{3}(\mu-1)}, 
\end{align} 
respectively, where  $\rho_{0\text{m}}$, $\rho_{0\text{r}}$, $a_0=1$ are the current values of the energy densities and the scale factor.
For $\mu=1$, the standard calculus is recovered, we have the standard evolution for CDM plus radiation, $\rho_{\text{m}}=\rho_{0\text{m}}a^{-3}$, $\rho_{\text{r}}=\rho_{0\text{r}}a^{-4}$ and \eqref{EH} becomes the standard Friedman equation in term of redshift, $E(z)^2  =\Omega_{0\text{m}}(z+1)^{3}
+\Omega_{0\text{r}}(z+1)^{4}$.

When $\mu\neq 1$, Eq. \eqref{EH} becomes
\begin{align}
  & E(z)^2+  (1-\mu) \frac{F(z)}{ t_U H_0}E(z)  \nonumber \\
    & =\Omega_{0\text{m}}(z+1)^{3}  F(z)^{(1-\mu)}
+\Omega_{0\text{r}}(z+1)^{4} F(z)^{\frac{4}{3}(1-\mu)}, \nonumber\\ \label{EH2}
\end{align}
where we have defined $\Omega_{0\text{m}}\equiv 8\pi G\rho_{0\text{m}}/3 H_0^2$, 
 $\Omega_{0\text{r}}\equiv 8\pi G\rho_{0\text{r}}/3 H_0^2$, $E(z) \equiv H(z)/H_0$ and $F(z)\equiv t_U/t(z)$. Note that $F(0)=1$ due to $t(0)=t_U$, is the age of the universe. 

Using the chain rule, we obtain a differential equation for $F(z)$ given by 
\begin{equation}
    F'(z)= \frac{dt}{dz}\frac{dF}{dt}= \frac{F^2(z)}{t_U H_0 (z+1)E(z)} \label{eqF}.
\end{equation}
When we solve \eqref{EH2} for $E(z)$, two branches are dictated by the sign $\pm$. The branch $-$ leads to $E\leq 0$ and the branch $+$ leads to $E \geq 0$. 

Therefore, since we are interested in an expanding universe, we choose the positive branch. That is, 
\begin{align}
     E(z)= -f F(z) 
   + F(z)^{-\mu } \Bigg\{ & f^2 F(z)^{2 (\mu +1)}   +\Omega_{0\text{m}} (z+1)^3
   F(z)^{\mu +1}  \nonumber \\
   & +\Omega_{0\text{r}} (z+1)^4 F(z)^{\frac{2 (\mu +2)}{3}}\Bigg\}^{1/2}, \label{FriedmannFinal1}
\end{align}
where $f\equiv(1-\mu)/(2 t_UH_0)$ is going to be the \textit{fractional constant} that will act as the cosmological constant.
Friedmann constraint gives us $f=(\Omega_{0\text{m}}+\Omega_{0\text{r}}-1)/2$, such that  for $\mu <1$, $\Omega_{0\text{m}}+\Omega_{0\text{r}}>1$,  for $\mu >1$, $\Omega_{0\text{m}}+\Omega_{0\text{r}}<1$, and notice that we choose the positive branch in order to have $E(z)>0$ and where $\Omega_{0\text{r}}=2.469\times10^{-5}h^{-2}(1+0.2271N_{\text{eff}})$, where $N_{\text{eff}}=2.99\pm0.17$ \citep{Planck:2018}.
The condition $\Omega_{0\text{m}}+\Omega_{0\text{r}}>1$, can be produced in a closed FLRW  universe.

In Eq. \eqref{FriedmannFrac}, the term $(1-\mu)H t^{-1}$ contributes as a positive term for $\mu<1$ or a negative term for $\mu>1$. 
Substituting \eqref{FriedmannFinal1} in \eqref{eqF}, we obtain a differential equation
\begin{align}
    F'(z)  = &\frac{2 f F(z)^{\mu +2}}{(\mu -1) (z+1)}  \times 
     \Bigg\{f F(z)^{\mu +1} \nonumber \\
    -\Bigg[  & f^2 F(z)^{2 \mu
   +2}  +\Omega_{0\text{m}} (z+1)^3 F(z)^{\mu +1} \nonumber \\
     &  +\Omega_{0\text{r}} (z+1)^4 F(z)^{\frac{2 (\mu +2)}{3}}\Bigg]^{1/2}\Bigg\}^{-1}, \label{F'(z)}
\end{align}
That has to be solved numerically.  
Then, we numerically calculate \eqref{FriedmannFinal1} plugging back the numerical results for $F(z)$. 

Moreover, the deceleration parameter $q(z)$ can be written as 
\begin{align}
    q(z)= -1 +(1+z) \frac{d \ln E(z)}{d z}. \label{q(z)}
\end{align}
Hence, substituting \eqref{FriedmannFinal1}  in  \eqref{q(z)}, using \eqref{eqF} to replace $F'(z)$, and using \eqref{FriedmannFinal1} to eliminate the radical we obtain a closed form for $q(z)$ as given by  Eq. \eqref{DecelerationFinal}, which quantifies if the Universe is in an accelerated stage and under which conditions. 

Finally, the cosmographic parameter known as the {\it jerk}, which quantifies if the model tends to $\Lambda$ or its another kind of DE, can be written as
\begin{eqnarray}
    j=q(2q+1)+(1+z)\frac{dq}{dz}, \label{Jerk}
\end{eqnarray}
where $q$ is given by Eq. \eqref{DecelerationFinal}.

\subsection{Analytic solution to the fractional Friedmann equation} \label{FracFriedmann}
Notice that for $\mu\neq 1$,  the modified continuity equation \eqref{CE}, also yields  the condition 
\begin{align}
    \frac{8 \pi  G}{3} \sum_i p_i=\frac{2 (\mu -3) H}{t}+H^2-\frac{ (\mu -2) (\mu -1)}{t^2}.
\end{align}
Combining with \eqref{IntegrableH}
and \eqref{FriedmannFrac}, one obtains
\begin{equation}
   \dot{H}+\frac{2 (\mu -4) H }{t}+3 H ^2-\frac{(\mu -2) (\mu -1)}{t^2}=0, \label{NewIntegrableH}
\end{equation}
whose analytical solution is 
(see an analogous case in \citet{Shchigolev:2012rp}, Eq. (36))
\begin{equation}
 H(t)=   \frac{9-2 \mu }{6 t}+\frac{\sqrt{8 \mu  (2 \mu -9)+105} \left(1-\frac{2 c_1}{t^{\sqrt{8 \mu  (2 \mu
   -9)+105}}+c_1}\right)}{6 t}, \label{solution}
\end{equation} where 
\begin{equation} \label{eq:c1_newH}
    c_1=\frac{t_U^{\sqrt{8 \mu  (2 \mu -9)+105}} \left(-6 H_0 t_U-2 \mu +\sqrt{8 \mu  (2 \mu
  -9)+105}+9\right)}{6 H_0 t_U+2 \mu +\sqrt{8 \mu  (2 \mu -9)+105}-9},
\end{equation} 
is an integration constant depending on $\mu$, the $H_0$ value and the Universe's age, $t_U$.
 The relation between redshift $z$ and cosmic time $t$ is through the scale factor, 
\begin{align}
a(z) & :=(1+z)^{-1}\nonumber \\
& = \left[\frac{t^{\sqrt{8 \mu  (2 \mu
   -9)+105}}+c_1}{{t_U^{\sqrt{8 \mu  (2
   \mu -9)+105}}+c_1}}\right]^{\frac{1}{3}} \left[ \frac{t}{t_U}\right]^{\frac{1}{6} \left(-2 \mu -\sqrt{8 \mu  (2 \mu -9)+105}+9\right)}. \label{tz}
\end{align}
That also leads asymptotically to power-law scale factors for large $t$, having
\begin{equation}
    a(t)\simeq   t^{\frac{1}{6} \left(-2 \mu +\sqrt{8 \mu  (2 \mu -9)+105}+9\right)}, \label{H(t)}
\end{equation}
  For $\mu \notin\{1,2\}$ and for large $t$, we acquire $q<0$, and then we have late-time acceleration without DE. The respective $E(t)$ and $q(t)$ parameters are shown in Appendix \ref{AlternativeApp}.

\section{Methodology and dataset} \label{sec:constraints}

A Bayesian Markov Chain Monte Carlo (MCMC) analysis is performed to constrain the phase-space parameter ${\bm\Theta}=\{h, \Omega_{0m}, \mu\}$ of the fractional cosmology using observational Hubble data OHD, SnIa dataset and joint analysis. Under the \texttt{emcee} Python package environment \citep{Foreman:2013}, after the auto-correlation time criterion warranty the convergence of the chains, a set of 4000 chains with 250 steps each is performed to establish the parameter bounds. Additionally, the configuration for the priors are Uniform distributions allowing vary the parameters in the range $h\in[0.2, 1]$,  $\Omega_{0m}\in[0,1]$ and $\mu \in[1,3]$.
Hence the figure-of-merit for the joint analysis is built through the Gaussian log-likelihood given as $-2\ln(\mathcal{L}_{\rm data})\varpropto \chi^2_{\rm data}$ and
\begin{equation} \label{eq:chi2_joint}
\chi^2_{\rm Joint} = \chi_{\rm CC}^2 + \chi_{\rm SnIa}^2 \,,
\end{equation}
where each term refers to the $\chi^2$-function for each dataset. Now, each piece of data is described in the rest of the Section.

\subsection{Cosmic chronometers}

Up to now, a set of 31 points obtained by differential age tools, namely cosmic chronometers (CC), represents the measurements of the Hubble parameter, which is cosmological independent \citep{Moresco:2016mzx}.
In this sense, this sample is useful to bound alternative models to $\Lambda$CDM. Thus, the figure-of-merit function to minimize is given by
\begin{equation} \label{eq:chiOHD}
    \chi^2_{{\rm CC}}=\sum_{i=1}^{31}\left(\frac{H_{th}(z_i)-H_{obs}(z_i)}{\sigma^i_{obs}}\right)^2,
\end{equation}
where the sum runs over the whole sample, and $H_{th}-H_{obs}$ is the difference between the theoretical and observational Hubble parameter at the redshift $z_i$ and $\sigma_{obs}$ is the uncertainty of $H_{obs}$.

\subsection{Type Ia Supernovae }

Ref. \citep{Scolnic:2018} provides 1048 luminosity modulus measurements, known as Pantheon sample, from Type Ia Supernovae which cover a region $0.01<z<2.3$. Due to this sample, the measurements are correlated, and it is convenient to build the chi-square function as
\begin{equation}\label{eq:chi2SnIa}
    \chi_{\rm SnIa}^{2}=a +\log \left( \frac{e}{2\pi} \right)-\frac{b^{2}}{e},
\end{equation}
where
\begin{eqnarray}
    a &=& \Delta\boldsymbol{\tilde{\mu}}^{T}\cdot\mathbf{Cov_{P}^{-1}}\cdot\Delta\boldsymbol{\tilde{\mu}}, \nonumber\\
    b &=& \Delta\boldsymbol{\tilde{\mu}}^{T}\cdot\mathbf{Cov_{P}^{-1}}\cdot\Delta\mathbf{1}, \\
    e &=& \Delta\mathbf{1}^{T}\cdot\mathbf{Cov_{P}^{-1}}\cdot\Delta\mathbf{1}, \nonumber
\end{eqnarray}
and $\Delta\boldsymbol{\tilde{\mu}}$ is the vector of residuals between the theoretical distance modulus and the observed one, $\Delta\mathbf{1}=(1,1,\dots,1)^T$, $\mathbf{Cov_{P}}$ is the covariance matrix formed by adding the systematic and statistic uncertainties, i.e.   $\mathbf{Cov_{P}}=\mathbf{Cov_{P,sys}}+\mathbf{Cov_{P,stat}}$. The super-index $T$ on the above expressions denotes the transpose of the vectors.

The theoretical distance modulus is estimated by
\begin{equation}
    m_{th}=\mathcal{M}+5\log_{10}\left[\frac{d_L(z)}{10\, pc}\right],
\end{equation}
where $\mathcal{M}$ is a nuisance parameter which has been marginalized by Eq. \eqref{eq:chi2SnIa}. 

The luminosity distance, denoted as $d_L(z)$, is computed through  
\begin{equation}\label{eq:dL}
    d_L(z)=(1+z)c\int_0^z\frac{dz^{\prime}}{H(z^{\prime})},
\end{equation}
being $c$ the speed of light.

\section{Results} \label{Results}

Table \ref{tab:bestfits} presents the cosmological constraints of fractional cosmology for CC and SnIa, samples
and the joint analysis, respectively. Each best-fit parameter value includes uncertainty at 68\% confidence level (CL).

\begin{table*}
\centering
\begin{tabular}{|lcccc|}
\hline
Sample &    $\chi^2_{\rm min}$     &  $h$ & $\Omega_{0m}$ & $\mu$ \\
\hline
CC    & $16.14$ & $0.629^{+0.027}_{-0.027}$  & $0.399^{+0.093}_{-0.122}$  & $2.281^{+0.492}_{-0.433}$  \\ [0.9ex] 
SnIa  & $54.83$ & $0.599^{+0.275}_{-0.269}$  & $0.160^{+0.050}_{-0.072}$  & $2.771^{+0.161}_{-0.214}$  \\ [0.9ex]
Joint & $78.69$ & $0.692^{+0.019}_{-0.018}$  & $0.228^{+0.035}_{-0.040}$  & $2.839^{+0.117}_{-0.193}$  \\ [0.9ex] 
\hline
\end{tabular}
\caption{Best-fit values and their $68\%$ CL uncertainties for fractional cosmology with CC, SnIa and a Joint analysis.}
\label{tab:bestfits}
\end{table*}

\begin{figure*}
    \centering
    \includegraphics[width=0.5\textwidth]{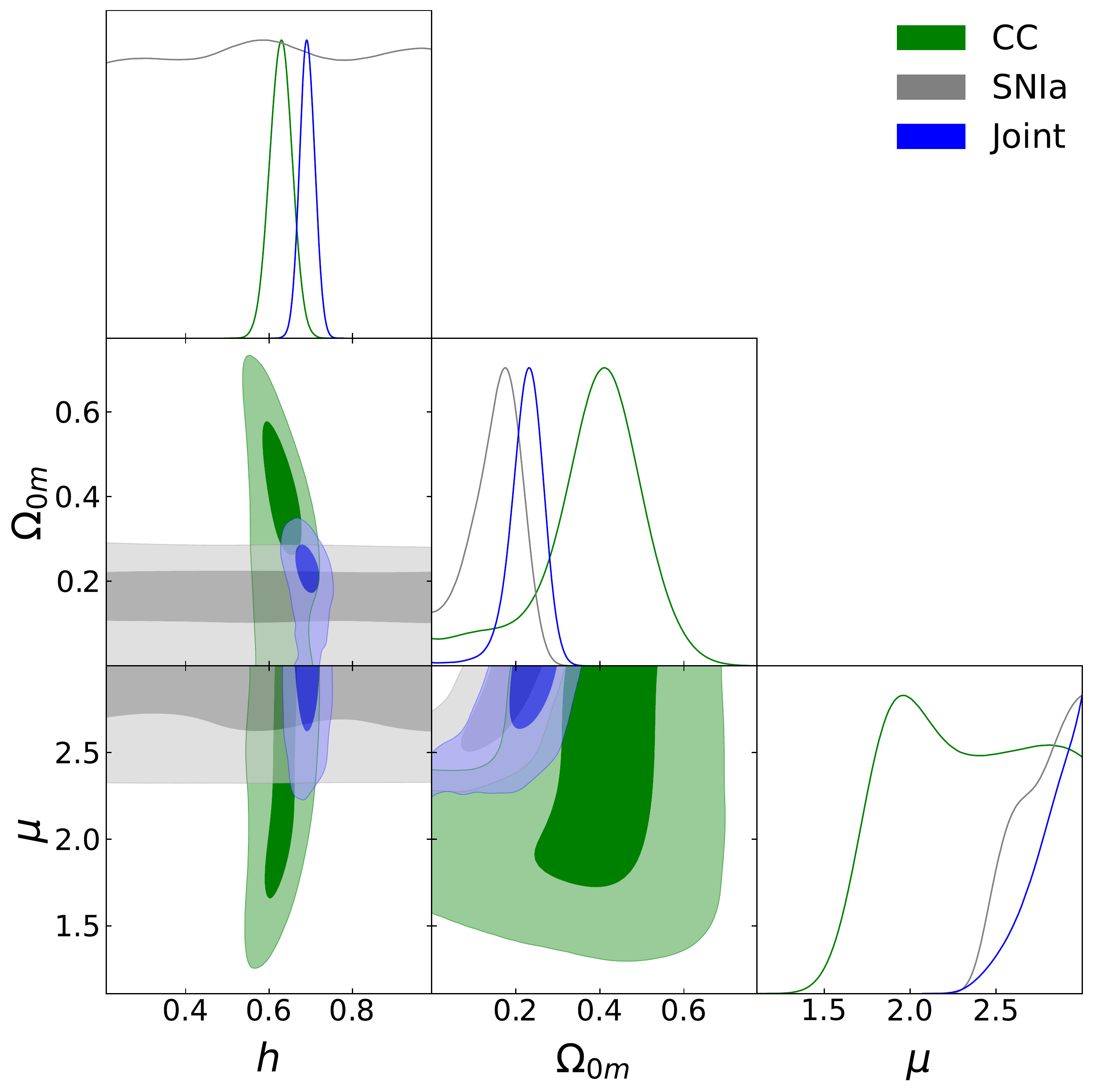}
    \caption{2D likelihood contours at 68\% and 99.7\% CL, alongside the corresponding 1D posterior distribution of the free parameters, in fractional cosmology.}
    \label{fig:contours}
\end{figure*}

Figure \ref{fig:contours} shows the 1D marginalized posterior distributions for each data and
joint analysis and also the 2D phase space distribution
at 68\% ($1\sigma$), 99.7\% ($3\sigma$) CL. According to the $\chi^2$ value, the model is in good agreement with the data. Furthermore, the characteristic parameter of the fractional cosmology, $\mu$, is estimated for each dataset, and in particular, we have $\mu=2.8393^{+0.117}_{-0.193}$ for the joint analysis, allowing an accelerated Universe. Notice that we recover traditional calculus when $\mu=1$; however, in the region $0<\mu<1$, obtaining an accelerated physical Universe (with non-negative ``age'') at late stages is not feasible. For this $\mu$-range, we can obtain an accelerated power-law solution corresponding to negative values for the age of the Universe; thus, the corresponding solution is nonphysical. Hence, one way to avoid this affliction is under the introduction of $\Lambda$, which will act as a cosmological constant. However, we are in a loop because the idea explains the Universe's acceleration through the $\mu$ term, which contributes to the fractional calculus theory. The other way is to consider $\mu>2$, and then we get an accelerated physical Universe at late stages.

\begin{figure*}
\centering
\includegraphics[width=0.32\textwidth]{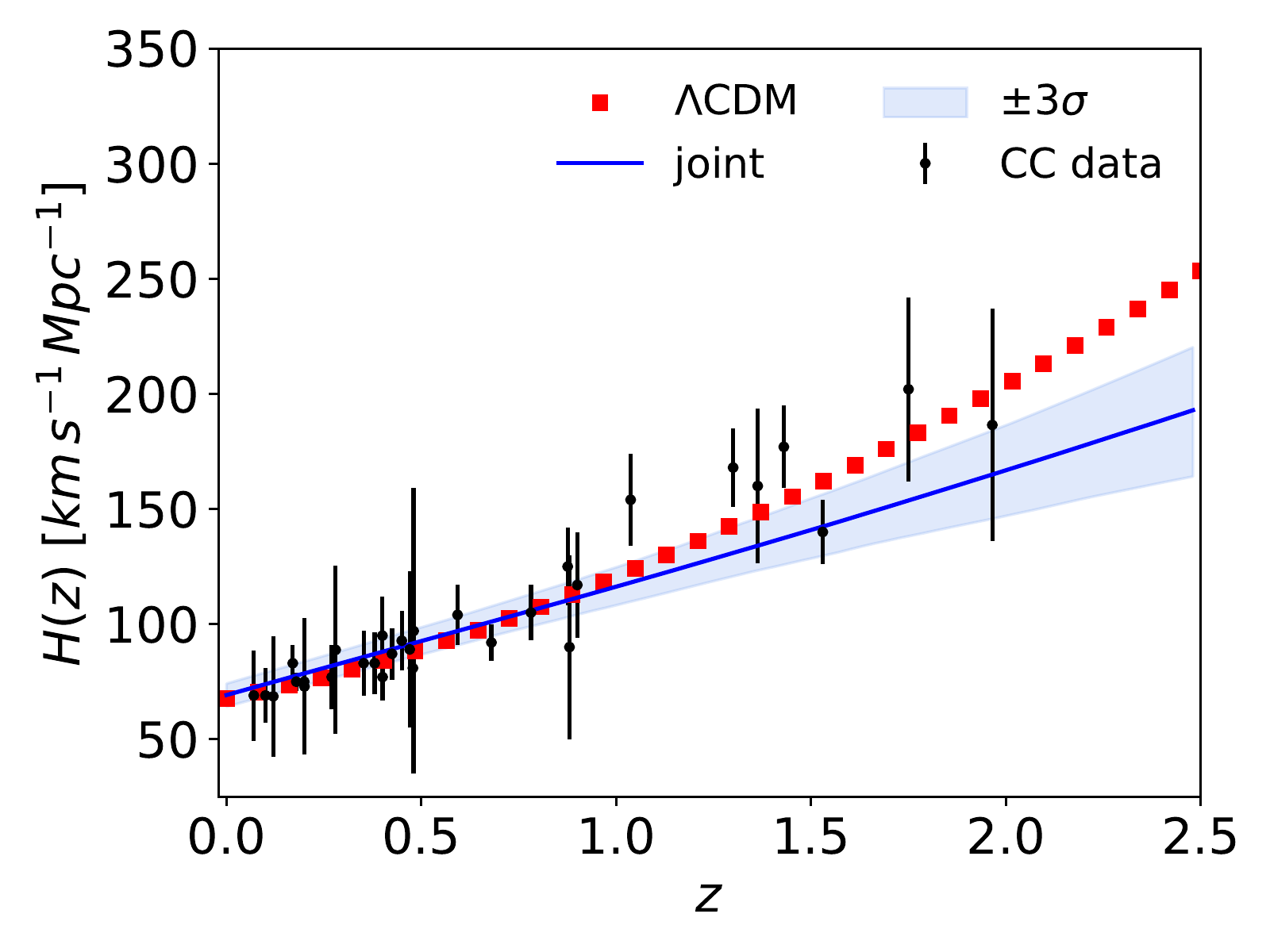}
\includegraphics[width=0.32\textwidth]{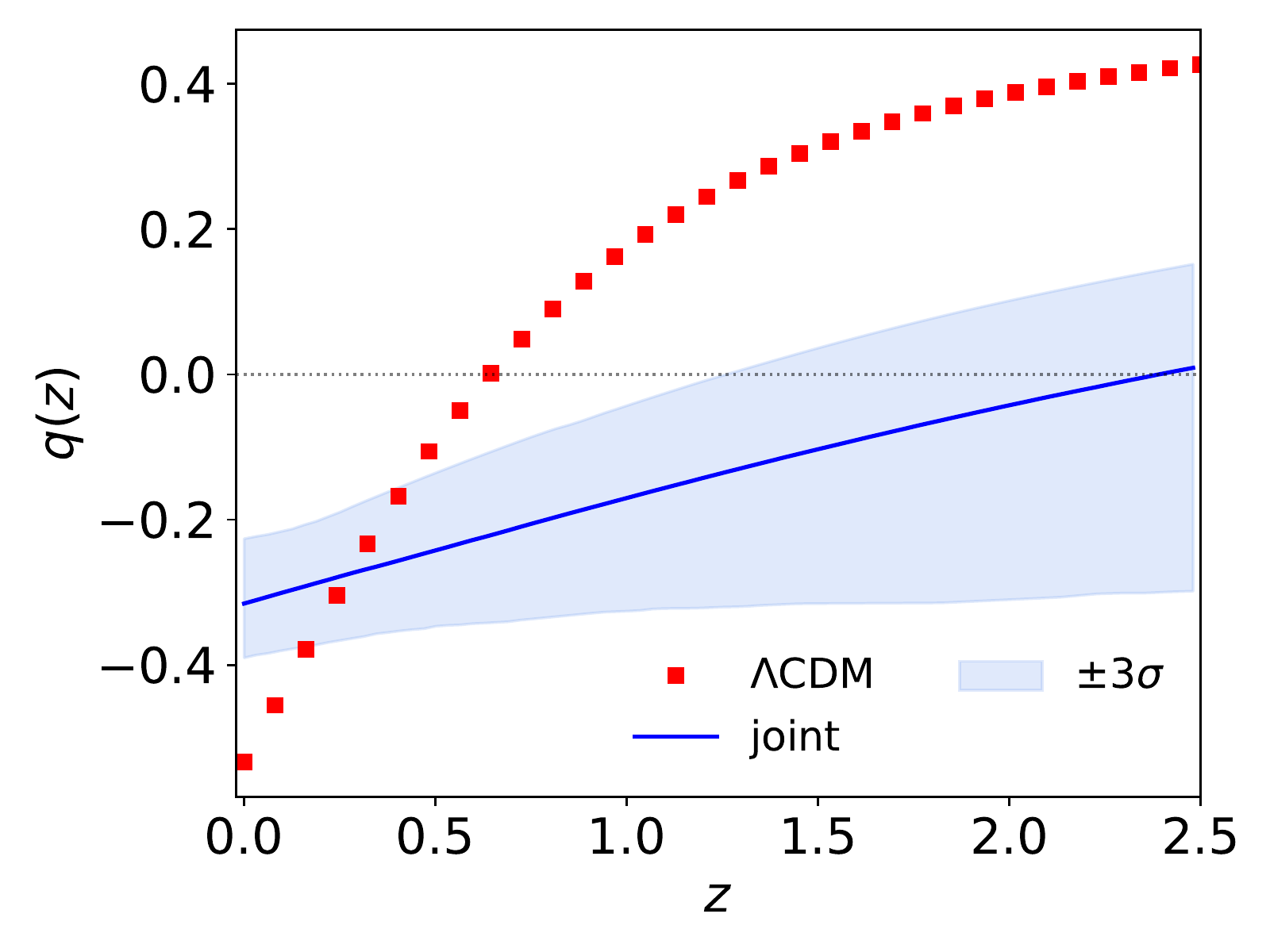}
\includegraphics[width=0.32\textwidth]{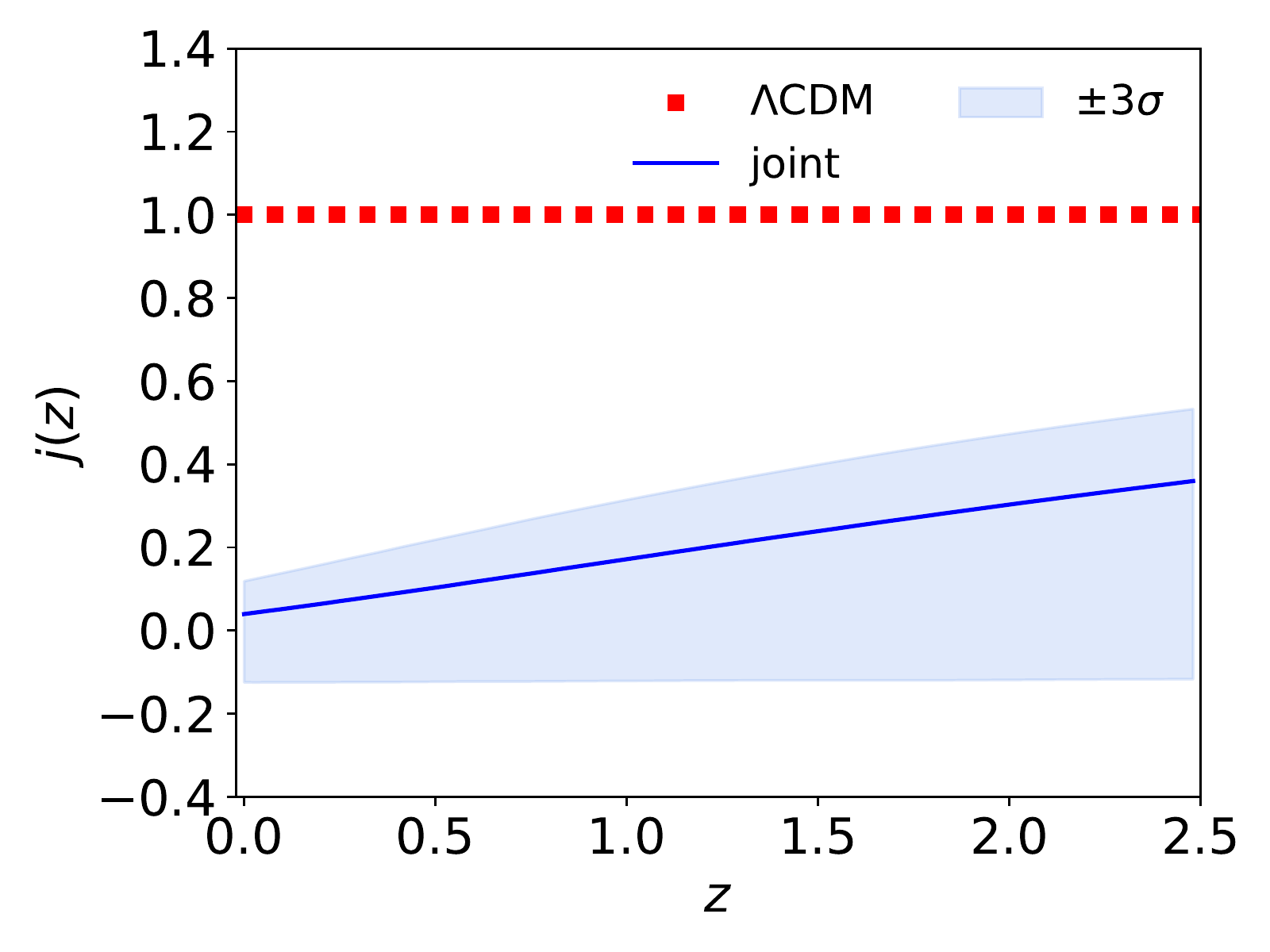}
\caption{Left to right: reconstruction of the $H(z)$, 
$q(z)$, 
and $j(z)$, 
in fractional cosmology represent the results of  $\Lambda$CDM cosmology  with $h=0.6766$ and
$\Omega_{\text{m}0}=0.3111$ 
\citep{Planck:2018}.}
\label{fig:Hz_and_qz}
\end{figure*}

On the other hand, the age of the Universe is estimated for each dataset, $t_U/{\rm Gyrs} = 33.633^{+14.745}_{-15.095}$ (CC), $33.837^{+27.833}_{-10.788}$ (SnIa) and $33.617^{+3.411}_{-4.511}$ (joint). For the Joint value, we obtain around $2.4$ times larger than the age of the Universe expected under the standard paradigm, which is also in disagreement with the value obtained with globular clusters, $t_U=13.5^{+0.16}_{-0.14}\pm 0.23$ \citep{Valcin:2021}. The term $(1-\mu)H/t$, which acts as an extra source of mass leading to a closed Universe, could be the origin of this older Universe. In closed scenarios, the Universe becomes older than the standard prediction \citep{DiValentino:2020srs}.

Regarding the cosmographic parameters at $z=0$ we have $q_0=-0.315^{+0.030}_{-0.028}$ and $j_0=0.040^{+0.032}_{-0.053}$ using the joint analysis. Furthermore, the redshift transition between acceleration and deceleration stages of $z_T=2.388^{+0.610}_{-0.510}$ is estimated. From Figure \ref{fig:Hz_and_qz}, the $z_T$, $q_0$ and $j_0$ values for fractional cosmology are deviated more than $3\sigma$ to the value obtained by $\Lambda$CDM. In contrast to $\Lambda$CDM, the reconstruction of a jerk for the alternative cosmology suggests an effective dynamical equation of state for the Universe for late times.

\begin{figure}
    \centering
    \includegraphics[width=0.45\textwidth]{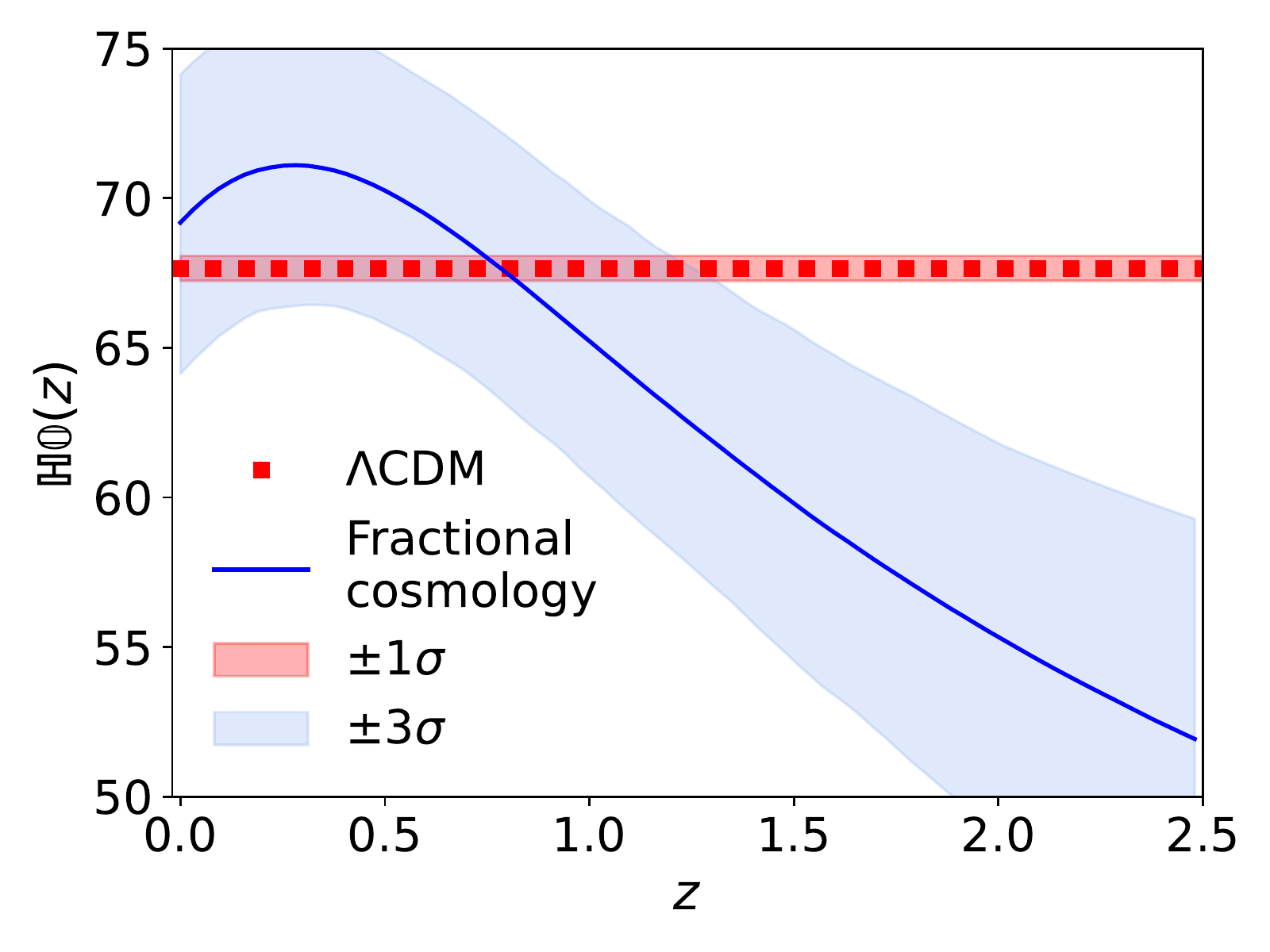}
    \caption{$\mathbf{\mathbb{H}}0(z)$ diagnostic for fractional cosmology and its comparison against $\Lambda$CDM model.}
    \label{fig:H0diagnostic}
\end{figure}

Figure \ref{fig:H0diagnostic} displays the reconstruction of the $\mathbf{\mathbb{H}}0(z)$ diagnostic \citep{H0diagnostic:2021} for the fractional cosmology and its error band at $3\sigma$ CL. Although the path (solid line) for the fractional cosmology is consistent within $3\sigma$ with the CMB Planck value \citep{Planck:2018} for $z\lesssim1.5$, we can observe that it presents a trend to the $H_{0}$ value obtained by SH0ES \citep{Riess:2019cxk} for the present time. Nevertheless, the $H_0$ value for $1.5<z<2.5$ is lower than the Planck value, suggesting a tension in this value.

\section{Dynamical Systems and Stability Analysis} \label{DS}

Defining the dimensionless age parameter $A=t H$, the re-scaled (dimensionless) energy density $\varrho_i= t^2 \rho_i$ where $\rho_i$ is defined in \eqref{EQ16}, and the logarithmic time $\tau=\ln t$ such that for any function $g$ we have $dg/d\tau= t dg/dt$. 

For the new variables, we have a restriction
\begin{equation}
   A^2+ (1-\mu)A=\frac{8\pi G}{3}\sum_i \varrho_i, \label{FriedmannFrac2}
\end{equation}
and evolution equations for each species
\begin{equation}
  \frac{d\varrho_i}{d\tau}= \varrho_i \left(1+ \mu +(\mu -1) w_{i}  -3 A (w_{i}+1)\right). \label{Matter-cons}
\end{equation}
To compare with the standard model we impose that the universe components are two ($n=2$), CDM  ($\mu_{1}=\varrho_{\text{m}}, w_1=w_{\text{m}}=0$) and radiation  ($\mu_{2}=\varrho_{\text{r}}, w_2=w_{\text{r}}=1/3$), which in our modified scenario  evolve according to
 \begingroup\makeatletter\def\f@size{9}\check@mathfonts
\begin{align}
    & \frac{d\varrho_{\text{m}}}{d \tau}= \varrho_{\text{m}}  (\mu -3 A+1), \label{SYST1}
\\
    & \frac{d\varrho_{\text{r}}}{d \tau}= \frac{2 \varrho_{\text{r}}  (2 \mu -6 A +1)}{3}, \label{SYST2}
 \\
    & \frac{d A}{d\tau}=  \frac{8 \pi  G (3 (\mu -3 A+1) \varrho_{\text{m}} +2 (2 \mu -6 A+1) \varrho_{\text{r}})}{9 (1-\mu +2 A)}.\label{SYST3}
\end{align}
\endgroup
Now, we present a reduced phase space which is determined by a coupled system $d\mathbf{X}/d\tau = \mathbf{F}(\mathbf{X})$ subject to a constraint
$G(\mathbf{X}) = \mathbf{0}$ ($\mathbf{X}$ constitute the reduced phase space variables). Of central importance
to the investigation of the dynamical system are the equilibrium points which are determined by
the equations $\mathbf{F}(\mathbf{X}) = \mathbf{0}, \mathbf{G}(\mathbf{X}) = \mathbf{0}$. We calculate the gradient   $\nabla\mathbf{G}(\mathbf{X})$,
that is used to solve the constraint to linear order locally. Defining  the dimensionless variables 
\begin{equation}
    x_{1}= \frac{8\pi G \varrho_{\text{m}} }{3  \left(A+ \frac{|1-\mu|}{2}\right)^2}, \quad   x_{2}= \frac{8\pi G \varrho_{\text{r}} }{3  \left(A+ \frac{|1-\mu|}{2}\right)^2}, \label{eq32}
\end{equation}
that evolve as  
\begingroup\makeatletter\def\f@size{8}\check@mathfonts
\begin{align}
 \frac{d x_1}{d\tau} & =\frac{x_1}{3-3 \mu +6 A}  \big[-3 \mu ^2+6 A^2 (3 x_1+4 x_2-3) \nonumber \\
& +| 1-\mu |  (9 A x_1+2 x_2
   (-2 \mu +6 A-1)-3 (\mu +1) x_1) \nonumber \\
   & -A (-15 \mu +6 (\mu +1) x_1+(8 \mu +4)
   x_2+3)+3\big],  \label{syst1}   
\\
\frac{d x_2}{d\tau}& =\frac{x_2 }{3-3 \mu +6 A} \big\{2 \big[-2 \mu ^2+\mu +3 A^2 (3 x_1+4
   x_2-4)\nonumber \\ & -3 (\mu +1) A x_1 
     -2 A (-5 \mu +2 \mu  x_2+x_2+2)+1\big] \nonumber \\
   & +| 1-\mu |  (9
   A x_1+2 x_2 (-2 \mu +6 A-1)-3 (\mu +1) x_1)\big\}, \label{syst2}
\\
\frac{d A}{d\tau}& = -\frac{(2 A+|
   1-\mu | )^2 (9 A x_1+2 x_2 (-2 \mu +6 A-1)-3 (\mu +1) x_1)}{12 (-\mu +2
   A+1)}, \label{syst3}
\end{align}
\endgroup
subject to the restriction 
\begin{equation}
G(x_{1}, x_{2}, A):= A^2+ (1-\mu)A-\frac{1}{4} (x_1+x_2) (2 A+| 1-\mu | )^2=0. \label{syt4}
\end{equation}
As a plausible physical conditions we assume $0 \leq \Omega_{{\text{m}}}:=\frac{x_1 (2 A +| 1-\mu | )^2}{4 A^2}\leq 1,  x_2\geq 0, A \geq 0$. 
The physical parameter region we have considered is $1\leq\mu\leq3$. 

Generically, evaluated at a fixed point $P$, we have 
\begin{align*}
 &  \rho_{\text{m}}(t)=\frac{3 x_1 (2 A +| 1-\mu | )^2}{32 \pi  G t^2},   \rho_{\text{r}}(t)= \frac{3  x_2 (2 A +| 1-\mu | )^2}{32 \pi  G t^2},  H(t)= \frac{A}{t}.
\end{align*}
Therefore, we have two physical observables defined as the dimensionless densities
\begin{align}
 \Omega_{\text{m}}= \frac{x_1 (2 A +| 1-\mu | )^2}{4 A^2}, \quad \Omega_{\text{r}}=\frac{x_2 (2 A +| 1-\mu | )^2}{4 A^2}, 
\end{align}
and the deceleration parameter, which can be written as 
\begin{align}
    q = -1+\frac{1}{A}  -\frac{(2 A+| 1-\mu | )^2 (3 x_1 (\mu -3 A+1)+2 x_2 (2 \mu -6 A+1))}{12 A^2 (-\mu +2
   A+1)}. \label{eq-q-1}
\end{align}

\begin{table*} 
\resizebox{\textwidth}{!}{
\begin{tabular}{|c|c|c|c|c|c|}
\hline 
Label & $(x_1, x_2, A)$ &  Existence   & Eigenvalues & Stability & $\nabla G(x_{1}, x_{2}, A) |_P$ \\\hline
$P_1$& $\left(0, 0, 0\right)$ & $1\leq \mu \leq 3$  & $\left\{0,\mu +1,\frac{2}{3} (2 \mu +1)\right\}$ & Source & $\left(-\frac{| 1-\mu | ^2}{4}, -\frac{| 1-\mu | ^2}{4}, 1-\mu\right)$ \\\hline
$P_2$& $\left(0, -\frac{(2 \mu +1) (4 \mu -7)}{(2 \mu +1 +3 | 1-\mu |)^2}, \frac{1}{6} (2 \mu
   +1)\right)$ & $1\leq \mu \leq \frac{7}{4}$&$\left\{0,\frac{1}{2},-\frac{(2 \mu +1) (4 \mu -7)}{3 (\mu -4)}\right\}$ & Saddle & $\left( -\frac{1}{36} (2 \mu +1 +3 | 1-\mu |)^2, -\frac{1}{36} (2 \mu +1 +3 | 1-\mu |)^2, -\frac{-2 \mu
   ^2+\mu +(\mu -4) | 1-\mu | +1}{2 \mu +1 +3 | 1-\mu |}\right)$ \\\hline 
   $P_3$ & $\left( -\frac{8 (\mu -2) (\mu +1)}{(2+ 2 \mu +3 | 1-\mu |)^2}, 0,  \frac{\mu +1}{3}\right)$ & $1\leq \mu <\frac{5}{2}$ & $\left\{0,-\frac{2}{3},-\frac{2 (\mu -2) (\mu +1)}{\mu -5}\right\}$ & Sink & $\left( -\frac{1}{36} (2 \mu +2 +3 | 1-\mu |)^2, -\frac{1}{36} (2 \mu +2 +3 | 1-\mu |)^2, \frac{2 \left(\mu
   ^2-1\right)-(\mu -5) | 1-\mu | }{2 \mu +2 +3 | 1-\mu |}\right)$ \\\hline
$P_4$ & $(0,0,\mu -1)$ & $1 \leq \mu \leq 3$ & $\left\{0,2(2- \mu),\frac{2}{3} (7-4 \mu )\right\}$& Source for $\mu <\frac{7}{4}$  & \\
&&&& Saddle for $\frac{7}{4}<\mu <2$ & \\
&&&& Sink for $\mu >2$ &$\left(-\frac{1}{4} (2 \mu-2 +| 1-\mu |)^2, -\frac{1}{4} (2 \mu-2 +| 1-\mu |)^2, \mu - 1\right)$ \\ \hline    
\end{tabular}}
\caption{\label{Tab1} Equilibrium points of system \eqref{syst1}, \eqref{syst2} and \eqref{syst3}. The physical parameter region is $1\leq\mu \leq 3$. The existence condition is $0 \leq  \frac{x_1 (2 A +| 1-\mu | )^2}{4 A^2} \leq 1,  x_2\geq 0, A \geq 0$.}
\end{table*}

In Tab. \ref{Tab1} the equilibrium points of system \eqref{syst1}, \eqref{syst2} and \eqref{syst3} which satisfy the restriction \eqref{syt4} are given.

\begin{table*}
\begin{tabular}{|c|c|c|c|c|c|c|}
\hline 
Label & $ \Omega_{\text{m}}$ & $ \Omega_{\text{r}}$  & $H$ & q & Solution & $a(t)= \left(t/t_U\right)^{A}$ \\\hline
$P_1$& Indeterminate & Indeterminate  & $0$ & Indeterminate & Static universe & $a(t)=\text{constant}$ \\\hline
$P_2$& $0$ & $\frac{7-4 \mu}{2 \mu +1}$ & $\frac{2 \mu +1}{6 t}$ & $ -\frac{2 \mu -5}{2 \mu +1}$& Power-law (decelerated if $\mu<\frac{5}{2}$) & \\
&&&& &  Power-law (accelerated if $\mu>\frac{5}{2}$) &  $a(t)= \left(t/t_U\right)^{ (2 \mu
   +1)/6}$ \\\hline 
$P_3$ &  $\frac{2 (2-\mu)}{\mu +1}$ & $0$ & $\frac{\mu +1}{3 t}$ & $-\frac{\mu-2}{\mu +1}$&  Power-law (decelerated if $\mu<2$) &\\
&&&& &  Power-law (accelerated if $\mu>2$) & $a(t)= \left(t/t_U\right)^{\frac{1+\mu}{3}}$\\\hline
$P_4$ & $0$ & $0$ & $\frac{\mu -1}{t}$ & $-\frac{\mu -2}{\mu -1}$ & Power-law (accelerated if $\mu<1$  or $\mu>2$) & \\
&&&& &  Power-law (decelerated if $1<\mu <2$) & $a(t)= \left(t/t_U\right)^{\mu-1}$ \\\hline 
\end{tabular}
\caption{\label{Tab2} Equilibrium points of system \eqref{syst1}, \eqref{syst2} and \eqref{syst3}. The physical parameter region is $1\leq\mu \leq3$. 
}
\end{table*}

In table \ref{Tab2} are presented the asymptotic values of the cosmological parameter for the  equilibrium points of system \eqref{syst1}, \eqref{syst2} and \eqref{syst3}, and the asymptotic expression of the scale factor.

\begin{table*}
\begin{tabular}{|c|c|c|}
\hline 
  $(x_1, x_2,A)$ & Eigenvalues & Solution \\\hline
  $ \left(0,  0,  0\right)$ & $\{0., 3.839, 4.452\}$ & $\rho_{\text{m}}(t)\to 0, \rho_{\text{r}}(t)\to 0,H(t)\to 0$ \\\hline
  $\left(0,-0.195601, 1.113\right)$ & $\left\{0,\frac{1}{2}, 8.35181\right\}$ & $ \rho_{\text{m}}(t)\to 0, \rho_{\text{r}}(t)\to  -\frac{0.0964524}{G t^2}, H(t)\to \frac{1.113}{t}$ (nonphysical)\\\hline  
  $\left(-0.147996, 0., 1.27967\right)$ & $\left\{0,-\frac{2}{3},2.98095\right\}$ & $\rho_{\text{m}}(t)\to  -\frac{0.0854376}{G t^2}, \rho_{\text{r}}(t)\to 0.,H(t)\to \frac{1.27967}{t}$  (nonphysical)\\\hline
  $({0., 0.,   1.839})$ & $\{0., -1.678, -2.904\}$ & $\rho_{\text{m}}(t)\to 0., \rho_{\text{r}}(t)\to 0., H(t)\to \frac{1.839}{t}$ \\\hline
\end{tabular}
\caption{\label{Tab3} Cosmological solutions represented by equilibrium points for the best-fit value $\mu=2.839$. }
\end{table*}

For the best fit value $\mu= 2.839$, the late-time attractor, and the equilibrium points corresponding to the cosmological solutions summarized in table \ref{Tab3}.

Following \citep{Hewitt:1992sk, Nilsson:1995ah, Goliath:1998mx}, we solve the restriction locally around the equilibrium points. This formulation will enable us to achieve a good understanding of the global structure of
the reduced phase space. 

 One eigenvalue is always zero due to the restriction  \eqref{syt4}. 
The expression $G(x_{1}, x_{2}, A)=0$ defines a singular surface,
with 
\begingroup\makeatletter\def\f@size{9}\check@mathfonts
\begin{align}
  \nabla G(x_{1}, x_{2}, A) & = \Bigg(-\frac{1}{4} (2 A+| 1-\mu | )^2,-\frac{1}{4} (2 A+| 1-\mu | )^2, \nonumber \\
  & \;\;\;\;\;  2 A+1-\mu -(x_1+x_2) (2 A+|
   1-\mu | )\Bigg). 
\end{align}
\endgroup
Notice that the gradient is different from zero at each point $P_i$ if $\mu \neq 1$. Therefore we can solve locally the restriction for each point $P_1$, $P_2$ and $P_3$,  say for $x_2\geq 0$. 
Hence,
\begin{equation}
\label{Eq:42}
x_2=\frac{4 A (-\mu +A+1)}{(2 A+| 1-\mu | )^2}-x_1. 
\end{equation}
 Notice that replacing \eqref{Eq:42} in \eqref{eq-q-1}, we acquire 
\begin{align}
q & =  \frac{4 \mu ^2-5 \mu +6 A^2-13 \mu  A+13 A+1}{6 A^2-3 \mu  A+3 A} \nonumber \\
& -\frac{x_1 (-\mu +3 A+1) (2 A+|
   1-\mu | )^2}{12 A^2 (-\mu +2 A+1)}.
\end{align}
{The dimensionless energy densities of matter and radiation reduces to 
\begin{align}
    \Omega_{\text{m}} & =\frac{x_1 (2 A+| \mu -1| )^2}{4 A^2},
\\
    \Omega_{\text{r}} & =1-\frac{(\mu -1)^2 x_1}{4 A^2}-\frac{\mu +x_1 |
   \mu -1| -1}{A}-x_1.
\end{align}}

Moreover, in the physical parameter region is $1\leq\mu \leq 3$, we obtain the two-dimensional dynamical system
\begin{align}
\frac{d x_1}{d\tau}& =  \frac{1}{3} x_1  \Bigg[3 \mu +\frac{6 A (10 A-3)}{\mu +2 A-1}+\frac{2 A (2 A
   (x_1+1)-3)}{\mu -2 A-1} \nonumber \\
   & -A (x_1+25)-\mu  x_1+x_1+3\Bigg], \label{reducedx1}
\\
 \frac{d A}{d\tau} &=  
 \frac{1}{12 (2 A+1-\mu)}\Bigg[12 A^3 (x_1-4)   -8 A^2 (-\mu  (x_1+8)+x_1+5)
 \nonumber \\
 & -(\mu -1) A (16 \mu +(\mu -1)
   x_1+8)-(\mu -1)^3 x_1\Bigg], \label{reducedA}
\end{align}

\begin{figure*}
    \centering
    \includegraphics[scale=0.75]{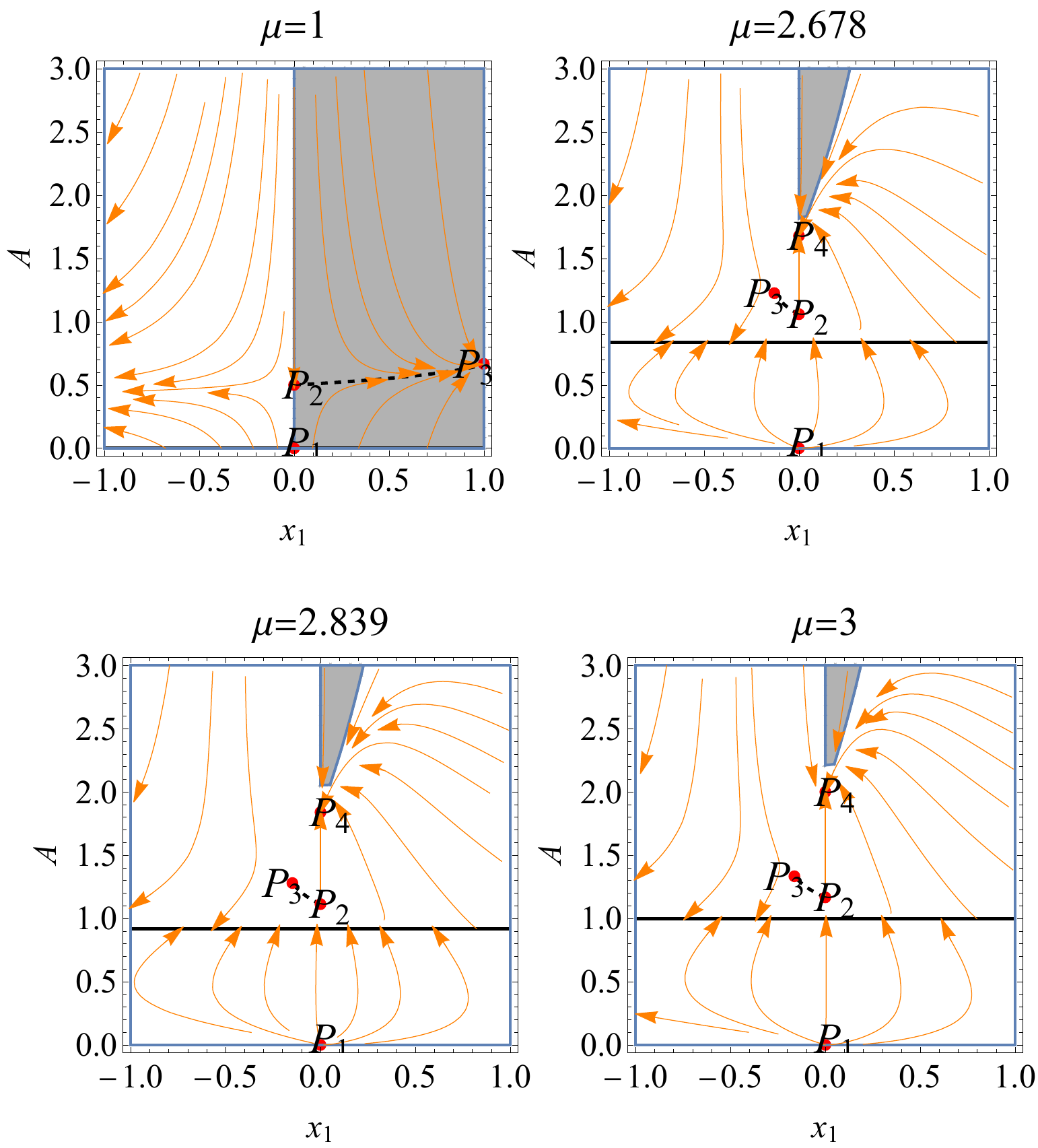}
    \caption{Phase flow of the reduced system \eqref{reducedx1} and \eqref{reducedA}. The shadowed region corresponds to $0\leq x_2\leq 1$.}
    \label{fig:DS1}
\end{figure*}

\begin{figure*}
    \centering
    \includegraphics[scale=0.75]{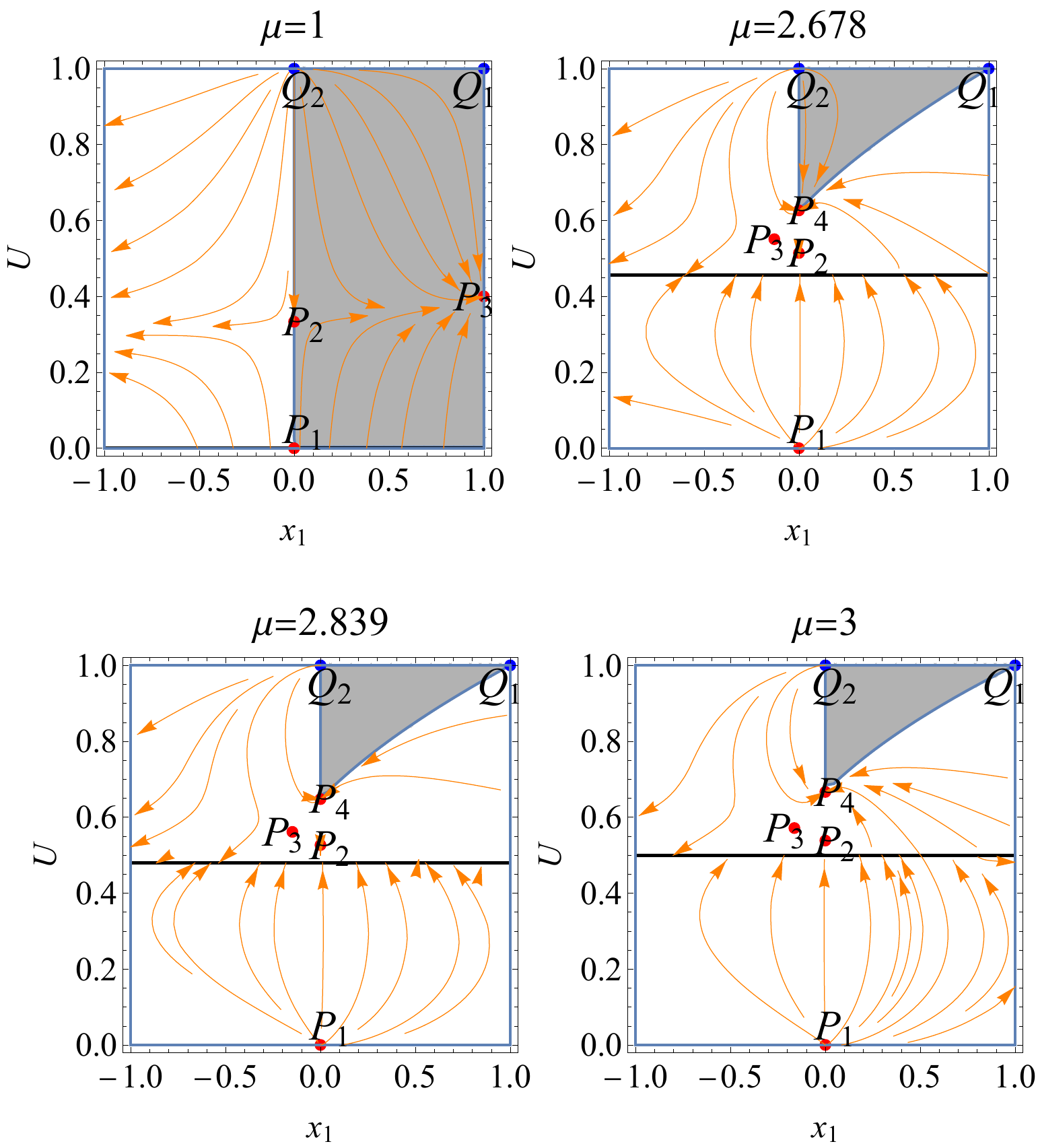}
    \caption{Phase flow of the reduced system \eqref{Preducedx1} and \eqref{PreducedA}. The shadowed region corresponds to $0\leq x_2\leq 1$.}
    \label{fig:DS2}
\end{figure*}

defined on the phase plane
\begin{eqnarray}
    & \Bigg\{(x_1, A)\in \mathbb{R}^2: 0\leq x_1 \leq 1, \; A \geq 0, \nonumber \\
    & 0\leq \frac{4 A (-\mu +A+1)}{(\mu +2 A-1)^2}-x_1\leq 1 \Bigg\}.
\end{eqnarray}
The equilibrium points of the reduced system are the same presented in Table \ref{Tab1}, where we now omit the zero eigenvalues. The singular line $A=(1-\mu)/2\leq 0$ is not on the physical region (the physical parameter region is $1\leq\mu \leq3$).

Now that we locally solved the constraint to linear order, the eigenvalues and eigenvectors
of the remaining locally unconstrained system are then listed.

\begin{enumerate}
    \item The eigensystem of $P_1: (x_1, A)=(0,0)$ (eigenvalues in first row; eigenvectors second row) is 
$\left(
\begin{array}{cc}
 \mu +1 & \frac{2}{3} (2 \mu +1) \\
 \left\{-\frac{4}{\mu -1},1\right\} & \{0,1\} \\
\end{array}
\right)$.

\item The eigensystem of $P_2: (x_1,A)=\left(0, \frac{1}{6} (2 \mu +1)\right)$ is 
$\left(
\begin{array}{cc}
 \frac{1}{2} & -\frac{(2 \mu +1) (4 \mu -7)}{3 (\mu -4)} \\
 \left\{-\frac{4 \left(16 \mu ^2-17 \mu -26\right)}{(5 \mu -2)^2},1\right\} & \{0,1\} \\
\end{array}
\right)$. 

\item The eigensystem of $P_3: (x_1,A)= \left(-\frac{8 (\mu -2) (\mu
   +1)}{(5 \mu -1)^2}, \frac{\mu +1}{3}\right)$ is 
$\left(
\begin{array}{cc}
 -\frac{2}{3} & -\frac{2 (\mu -2) (\mu +1)}{\mu -5} \\
 \left\{-\frac{12 (\mu -2) (\mu +1) (15 \mu -11)}{(5 \mu -1)^3},1\right\} & \left\{\frac{36 (\mu -1)
   (\mu +7)}{(5 \mu -1)^3},1\right\} \\
\end{array}
\right)$. 

\item The eigensystem of $P_4: (x_1,A)=(0, \mu-1)$ is 
$\left(
\begin{array}{cc}
 4-2 \mu  & \frac{2}{3} (7-4 \mu ) \\
 \left\{\frac{4}{9 (\mu -1)},1\right\} & \{0,1\} \\
\end{array}
\right)$.
\end{enumerate}
In Fig. \ref{fig:DS1} a phase flow of the reduced system \eqref{reducedx1} and \eqref{reducedA} is presented. The physical part of the phase plane, $x_2\geq 0$, is represented by a shaded region in the phase planes. It is confirmed that on the interval $1\leq \mu\leq 3$, $P_1$ is a source, $P_2$ is a saddle (it is nonphysical for $\mu>\frac{7}{4}$) and $P_3$ is a sink (it is nonphysical for $\mu>\frac{5}{2}$). $P_3$ satisfies $a(t)= \left(t/t_U\right)^{\frac{1+\mu}{3}}$, which is physical for  $1\leq \mu <\frac{5}{2}$. Evaluating the dimensionless energy densities $\Omega_{\text{m}}$, $\Omega_{\text{r}}$ of matter and radiation, and the Hubble parameter, we have $\Omega_{\text{m}}= -\frac{2 (\mu -2)}{\mu +1}$, $\Omega_{\text{r}}=0$, $H=\frac{\mu +1}{3 t}$ and  $q=-\frac{\mu -2}{\mu +1}$. It is a power-law (decelerated) late-time attractor for $\mu<2$.  Moreover, for $\mu>1$ the more interesting solution is $P_4$ that, as is presented in Figure \ref{fig:DS1}, it is the physical attractor for $\mu>2$. 
Moreover, the point $P_4$ satisfies  $a(t)= \left(t/t_U\right)^{\mu-1}$. It can be a source for $\mu <\frac{7}{4}$,  or a saddle for $\frac{7}{4}<\mu <2$ or a sink for $\mu >2$. This point does not exist in G.R. (for which $\mu=1$). The cosmological observable are $\Omega_{\text{m}}=0$, $\Omega_{\text{r}}=0$, $H=\frac{\mu -1}{t}$ and  $q=-\frac{\mu -2}{\mu -1}$. This solution is accelerated power-law if $\mu<1$ or $\mu>2$,  or decelerated power-law   if $1<\mu <2$. 

The previous system is not compact for $A$, so we define $U= {A}/(1+A)$ to obtain the dynamic system
\begin{align}
\frac{dx_1}{d\tau} =    \frac{1}{3} x_1 & \Bigg[(3 -x_1) (\mu-1)  +\frac{2 U (U (2 x_1+5)-3)}{U^2+\mu  (U-1)^2-1} \nonumber \\
& +\frac{6 U (13
   U-3)}{(U-1) (-\mu +(\mu -3) U+1)}+\frac{U (x_1+25)}{U-1}\Bigg], \label{Preducedx1}
\\
\frac{dU}{d\tau}=   \frac{1}{12}& \Bigg[\frac{8 U^2 (U (2 x_1+5)-3)}{\mu  (U-1)+U+1}+\mu ^2
   (U-1)^2 x_1 +x_1\nonumber \\
   & -\mu  (U-1) (U (5 x_1+16)-2 x_1) \nonumber \\
   & +U (2 U (x_1-20)-5
   x_1+8)\Bigg],\label{PreducedA}
\end{align}
defined on 
\begin{align}
  & \Bigg\{(x_1, U)\in [0,1]^2, 0 \leq \frac{4 U (\mu  (U-1)+1)}{(1-\mu +(\mu -3) U)^2}-x_1\leq 1\Bigg\}.
\end{align}

In Fig. \ref{fig:DS2} a phase flow of the reduced system \eqref{Preducedx1} and \eqref{PreducedA} is presented. Additionally to $P_1$, $P_2$, $P_3$ and  $P_4$, there appear two points at infinity. The point $Q_1: (x_1, U)= (1,1)$ that is a saddle and the point $Q_2: (x_1, U)= (0,1)$ that is a local source.

For the analysis of the unstable manifold of $P_2$, we consider the quantities $u, v$ defined in Appendix \ref{app2}, E.Q.s.  \eqref{B2}, and defines the graph $(u, g(u))$ in \eqref{B3} which satisfies the differential equation \eqref{ODE}. 
The cosmological solution associated to the unstable manifold of $P_2$ determines a curve in the physical space $(t^2 \rho_{\text{m}}, t^2 \rho_{\text{r}}, t H)$ given by 
\begin{align}
   & t^2 \rho_{\text{m}}(t)= -\frac{(\mu  (16 \mu -17)-26) u (5 \mu +6 g(u)+6 u-2)^2}{24 \pi  (2-5 \mu )^2  G},
\\
   &  t^2 \rho_{\text{r}}(t)= \frac{3 \left((2-5 \mu )^2+4 (\mu  (16 \mu -17)-26) u\right) \left(\frac{\mu
   }{3}+g(u)+u+\frac{1}{6}\right)^2}{8 \pi  (2-5 \mu )^2 G} \nonumber \\
   & +\frac{3 (1-\mu ) \left((2-5 \mu )^2+4 ((17-16
   \mu ) \mu +26) u\right) \left(\frac{\mu }{3}+g(u)+u+\frac{1}{6}\right)}{8 \pi  (2-5 \mu )^2 G} \nonumber \\
   & +\frac{3
   (\mu  (16 \mu -17)-26) (\mu -1)^2 u}{8 \pi  (2-5 \mu )^2 G},
\end{align}
\begin{align}
   & t H(t) =\frac{\mu }{3}+g(u)+u+\frac{1}{6}.
\end{align} such that 
\begin{align}
      & \Omega_{\text{m}} = -\frac{4 (\mu  (16 \mu -17)-26) u (5 \mu +6 g(u)+6 u-2)^2}{(2-5 \mu )^2 (2 \mu +6 g(u)+6    u+1)^2}, 
\\
   & \Omega_{\text{r}}= \frac{4 (\mu  (16 \mu -17)-26) u (5 \mu +6 g(u)+6 u-2)^2}{(2-5 \mu )^2 (2 \mu +6 g(u)+6
   u+1)^2} \nonumber \\
   & +\frac{6-6 \mu }{2 \mu +6 g(u)+6 u+1}+1.
\end{align}

\begin{figure}
    \centering
    \includegraphics[scale=0.8]{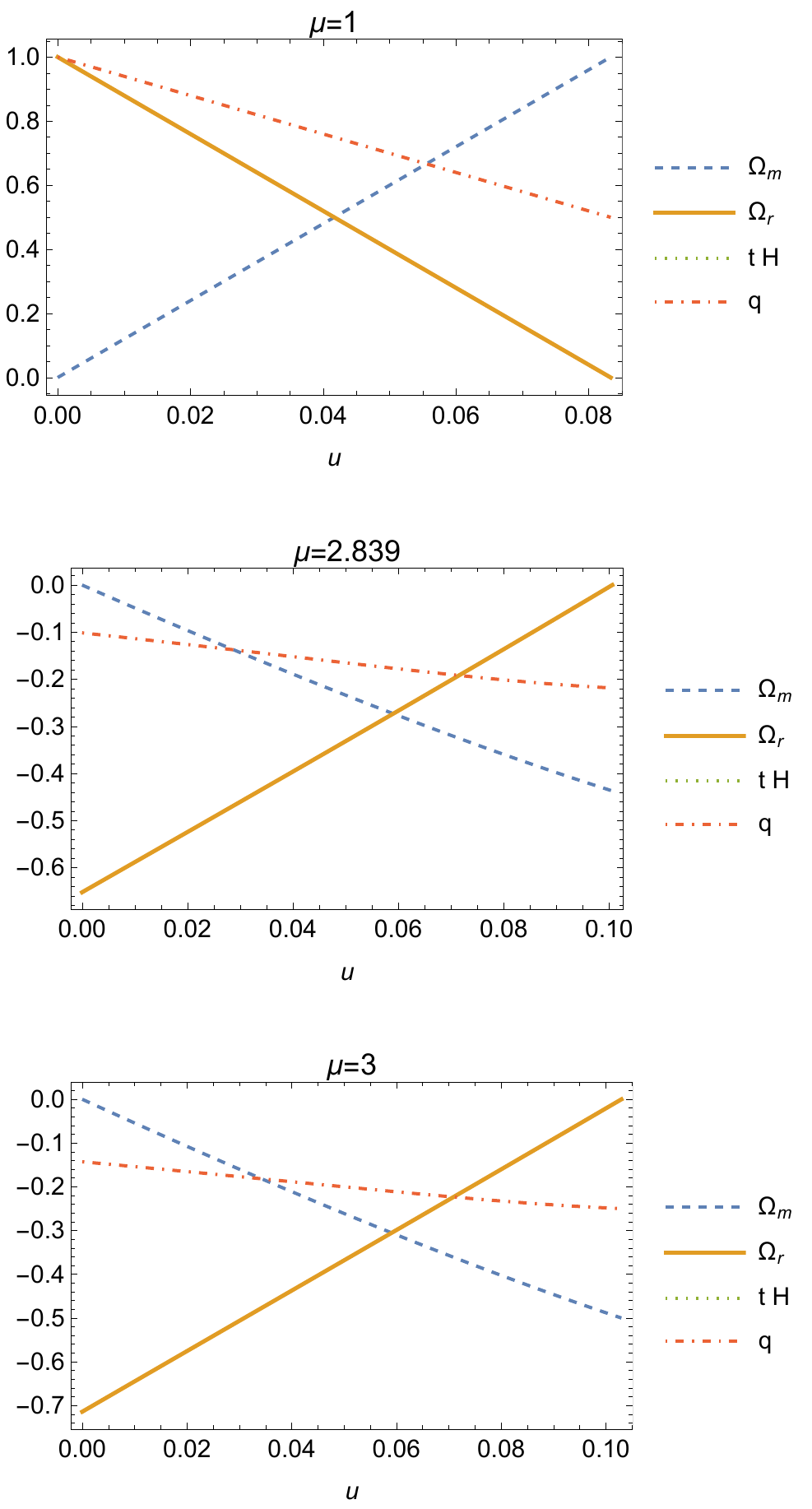}
    \caption{Evolution of $\Omega_{\text{m}}$, $\Omega_{\text{r}}, t H$ and $q$ vs $u$ for different values of $\mu$, as the flow moves along the unstable manifold connecting the saddle point $P_2$ with the sink $P_3$. For $\mu>7/4$ point  $P_2$ becomes nonphysical leading to negative $\Omega_{\text{r}}$, as well as $P_3$ which for $\mu>5/2$ leads to negative $\Omega_{\text{m}}$. $P_3$ becomes unstable for  $\mu>2$ emerging the late-time accelerated power-law solution $P_4$ for $\mu>2$. The middle and lower panels show that we have acceleration without Dark Energy.}
    \label{Fig3}
\end{figure}
In figure \ref{Fig3} is presented the evolution of $\Omega_{\text{m}}$, $\Omega_{\text{r}},t H$ and $q$ vs $u$ for the values $\mu\in \{1.0, 2.839, 3.0\}$. {Observe that as $\mu\approx 1$, at $P_3$, $\Omega_{\text{r}}\sim 0$, $\Omega_{\text{m}}\sim 1$, in complete analogy with the CDM model ($\Lambda=0$)}. For $\mu>7/4$ point  $P_2$ becomes nonphysical leading to negative $\Omega_{\text{r}}$, as well as $P_3$ which for $\mu>5/2$ leads to negative $\Omega_{\text{m}}$. $P_3$ becomes unstable for  $\mu>2$ emerging the late-time accelerated power-law solution $P_4$ for $\mu>2$.

\section{Bianchi I Universe} \label{Bianchi}
In the Misner variables, the L.R.S. Bianchi I spacetime is described by the line
element
\begin{equation}
ds^{2}=-N^{2}\left(  t\right)  dt^{2}+e^{2\alpha\left(  t\right)  }\left(
e^{2\beta\left(  t\right)  }dx^{2}+e^{-\beta\left(  t\right)  }\left(
dy^{2}+dz^{2}\right)  \right) , \label{ch.03}%
\end{equation}
where $\alpha$ is the scale factor for the three-dimensional
hypersurface and $\beta$ is the anisotropic parameter while $N$ is the lapse function. For $\beta   \rightarrow0$,
the line element (\ref{ch.03}) reduces to the spatially flat FLRW geometry.
The Lagrangian of GR, i.e. the Ricci scalar, is
calculated as
\begin{equation}
R=\frac{1}{N^{2}}\left(  6\ddot{\alpha}-6\dot{\alpha}\frac{\dot{N}}{N}%
+12\dot{\alpha}^{2}+\frac{3}{2}\dot{\beta}^{2}\right) . \label{ch.06}%
\end{equation}
The fractional effective action can be written in the form
\begin{small}
\begin{align}
&  S_{\text{eff}}  = \frac{1}{\Gamma(\alpha)}\int_0^t \Bigg[\frac{3}{8\pi G N^2(\tau)}\Bigg(\ddot{\alpha}(\tau)-\frac{\dot{\alpha}(\tau)\dot{N}(\tau)}{N(\tau)}%
+2\dot{\alpha}^{2}(\tau)+\frac{1}{4}\dot{\beta}^{2}(\tau)\Bigg) \nonumber\\ 
    & \;\;\;\;\;\;\;\;\;\;\;\;\;\;\;\;\;\;\;\;\;\;\;\;\;\; +e^{3\alpha(\tau)}\mathcal{L}_{\text{m}}\Bigg]  (t-\tau)^{\mu-1} N(\tau) d\tau.\label{LFCBI}
\end{align}
\end{small}
Varying the action \eqref{LFCBI} for $q_i\in \{N,\alpha, \beta\}$, and using the gauge $N=1$ after the variation, we obtain  from the Euler-Poisson equations,  the equations of motion
 \begin{align}
   & \dot{\alpha}^2  +\frac{(1-\mu) \dot{\alpha} }{t} -\frac{1}{4} \dot{\beta}^2 = \frac{8 \pi G}{3} \rho , 
  \\
   &\ddot{\alpha } +\frac{(1-\mu ) \dot{\alpha} }{t}+\frac{3}{2}{\dot{\alpha}}^2 +\frac{3}{8} \dot{\beta}^2  +\frac{(\mu -2) (\mu -1)}{2 t^2}=-4 \pi  G p ,
\\
   & \dot{\beta}  \left(3 \dot{\alpha} +\frac{1-\mu }{t}\right)+\ddot{\beta} =0,
 \end{align}
where  $\rho= \sum_i\rho_i$ and  $p =\sum_i p_i$ denotes the total energy density and total pressures of the matter fields. Now $H=\dot{\alpha}$ and $\sigma=\dot{\beta}/2$ are respectively the Hubble parameter and the anisotropy parameter.

Therefore, the field equations can alternatively be written as  \begin{align}
   & H^2 +\frac{(1-\mu) H}{t} -\sigma^2= \frac{8 \pi G}{3} \sum_i\rho_i, \label{BIb}\\   
   &  \dot{H}+\frac{(1-\mu) H}{t}+\frac{3}{2}{H}^2 + \frac{(\mu -2) (\mu -1)}{2 t^2}+3 \sigma^2   = - 4 \pi  G  \sum_i  p_i , \label{BIa}\\
   & \dot{\sigma}+ 3\sigma \left(H+\frac{1-\mu }{3 t}\right)=0,
 \end{align}
 and we consider separated conserved equations
 \begin{equation}
     \dot{\rho}_i+3\left(H+\frac{1-\mu}{3t}\right)(\rho_i+p_i)=0. \label{cons2}
 \end{equation}
As before, it is expected that the continuity equation for a perfect fluid is the energy conservation law for the matter, which is followed using the Bianchi identity. 
For $\alpha \neq 1$, \eqref{cons2} also yields modified continuity equations, only if  
\begin{align}
    \frac{8 \pi  G}{3} \sum_i p_i=\frac{2 (\mu -3) H}{t}+H^2-\frac{(\mu -2) (\mu -1)}{t^2}-\sigma^2.\label{BIc}
\end{align}
Eliminating  $\sum_i p_i$ and  $\sum_i \rho_i$ from \eqref{BIb}, \eqref{BIa} and  \eqref{BIc}, results in the cancellation of the $\sigma$-terms. Therefore, obtaining the master equation \eqref{NewIntegrableH} that has the solution \eqref{solution} where $c_1$ is an integration constant depending of $\mu$, the value of $H$ today, $H_0$ and the age of the Universe, $t_U$. That also leads  to  $a(t)\simeq   t^{\frac{1}{6} \left(-2 \mu +\sqrt{8 \mu  (2 \mu -9)+105}+9\right)}$ for large $t$. 
  For $\mu \notin\{1,2\}$ and for large $t$, we acquire $q<0$, and then we have late-time acceleration without DE.

\subsection{Dynamical Systems and Stability Analysis}

Using the equation of state $p_i=w_i\rho_i$, where $w_i \neq -1$ are constants, and 
\begin{table*}
    \centering
     \resizebox{\textwidth}{!}{   
     \begin{tabular}{|c|c|c|c|c|c|c|}
    \hline
Label &  $\Omega _{\text{m}}$& $\Omega _{\text{r}}$ & $\Sigma$ & $A$ & $q$ & Stability \\\hline
$A$ &  $0$ & $0$ & $-1$ & \text{Infinity} & $2$ & Source for $1<\mu <3$ \\\hline
$B$ &  $0$ & $0$ & $0$ & $\mu -1$ & $\frac{3 \mu -8}{\mu -1}$ & Sink for $1<\mu <\frac{8}{3}$ \\
 &&&&&& Source for  $\mu >\frac{7}{2}$ \\
 &&&&&& Saddle for $\frac{8}{3}<\mu <\frac{23}{8}$ \\
 &&&&&& or $\frac{23}{8}<\mu <\frac{7}{2}$
 \\
 \hline
$C$ &  $0$ & $-\frac{\mu +\sqrt{\mu  (25 \mu -131)+187}-7}{3 (\mu -2)}$ & $0$ & $\frac{1}{3} \left(-4 \mu +\sqrt{\mu  (25 \mu
   -131)+187}+13\right)$  & $\frac{-2 \sqrt{25 \mu ^2-131 \mu +187}+\mu +8}{9 (\mu -2)}$ & Saddle \\\hline
$D$ &  $0$ & $\frac{-\mu +\sqrt{\mu  (25 \mu -131)+187}+7}{3 (\mu -2)}$ &$ 0$ & $\frac{1}{3} \left(-4 \mu -\sqrt{\mu  (25 \mu
   -131)+187}+13\right)$ & $\frac{2 \sqrt{25 \mu ^2-131 \mu +187}+\mu +8}{9 (\mu -2)}$ & Saddle \\\hline
$E$ & $-\frac{\mu +\sqrt{\mu  (49 \mu -242)+337}-9}{4 (\mu -2)}$ & $0$ & $0$ &$ \frac{1}{6} \left(-5 \mu +\sqrt{\mu  (49 \mu
   -242)+337}+17\right)$ & $-\frac{\sqrt{49 \mu ^2-242 \mu +337}+\mu -9}{8 (\mu -2)}$ & Sink for $\mu >\frac{8}{3}$  \\
   &&&&&&  Saddle for $1<\mu <2$ \\
   &&&&&&  or $2<\mu <\frac{8}{3}$\\\hline
$F$ & $ \frac{-\mu +\sqrt{\mu  (49 \mu -242)+337}+9}{4 (\mu -2)}$ & $0$ &$ 0$ & $\frac{1}{6} \left(-5 \mu -\sqrt{\mu  (49 \mu
   -242)+337}+17\right)$ & $\frac{\sqrt{49 \mu ^2-242 \mu +337}-\mu +9}{8 (\mu -2)}$ & Sink for $\mu >2$ \\
   &&&&&& Source for $1<\mu <2$\\\hline
$G $& $0$ & $0$ & $1$ & \text{Infinity} & $2$ & Source for $1<\mu <3$\\\hline
$H$ & $0$ & $0$ & $-\frac{\sqrt{7-2 \mu }}{\sqrt{\mu -2}}$ & $\frac{(\mu -2) (\mu -1)}{3 (\mu -3)}$& $-\frac{\mu -5}{\mu -2}$  & Saddle \\\hline
$I$ & $0 $& $0$ & $\frac{\sqrt{7-2 \mu }}{\sqrt{\mu -2}}$ & $\frac{(\mu -2) (\mu -1)}{3 (\mu -3)}$& $-\frac{\mu -5}{\mu -2}$ & Saddle \\\hline
    \end{tabular}}
    \caption{Equilibrium points of the system  \eqref{syst-1}, \eqref{syst-2} and  \eqref{syst-3}. We assume $1\leq \mu \leq 3$. The eigenvalues are summarized in table \ref{tab:eigenvalues}.}
    \label{tab:my_label}
\end{table*}
defining dimensionless variables 
\begin{equation}
    \Omega_i=  \frac{8 \pi G \rho_i}{3 H^2}, \quad \Sigma=\frac{\sigma}{H}, \quad A=t H,
\end{equation}
which satisfies
\begin{equation}   1 -\frac{(\mu -1)}{A}= \Sigma^2 + \sum_i\Omega_i \label{NewFried},
\end{equation}
and taking the new derivative $f^{\prime}=\dot{f}/H$, we obtain for $\mu \neq 1$, 
\begin{align}
\Omega_j^{\prime}& =\Omega_j \left[
   (2 q-3 w_j-1) +  (w_j+1)   \left(1- \Sigma^2 - \sum_i\Omega_i\right)\right],\\
  \Sigma^{\prime} &=\Sigma \left[ (q-2)  + \left(1- \Sigma^2 - \sum_i\Omega_i\right)\right], \\
 A^{\prime} & =  1- A (1+q), 
\end{align}
where the deceleration parameter is found from Eq. \eqref{NewIntegrableH} (valid for FLRW and Bianch I metrics) as 
\begin{equation}
q:= -1 -   \frac{\dot{H}}{H^2}= 2 + \frac{2 (\mu -4) }{A}-\frac{(\mu -2) (\mu -1)}{A^2}.
\end{equation}
Assuming $\mu \neq 1$ and using \eqref{NewFried} to eliminate the $A$, we have 
\begin{small}
\begin{align}
q = 2-\frac{(\mu -2) }{ (\mu -1)}\left(1- \Sigma^2 - \sum_i\Omega_i\right)^2+\frac{2 (\mu -4)}{ (\mu -1)}\left(1- \Sigma^2 - \sum_i\Omega_i\right),
\end{align}
\end{small} 

\begin{table*}
    \centering
    \resizebox{\textwidth}{!}{
    \begin{tabular}{|c|c|c|c|}\hline
  Label &  $\lambda_1$ &  $\lambda_2$ & $\lambda_2$\\\hline
$A $&      $  \frac{12}{\mu -1}-6$ & $2$ & $3$ \\\hline
$B$ & $2-\frac{5}{\mu -1}$ & $6-\frac{10}{\mu -1}$ & $\frac{16}{3}-\frac{10}{\mu -1}$ \\\hline
$C$&  $\frac{\sqrt{25 \mu ^2-131 \mu +187}-5 \mu +5}{18-9 \mu }$ &$ -\frac{4 \left(\mu  \left(25 \mu +\sqrt{\mu  (25 \mu
   -131)+187}-131\right)-7 \sqrt{\mu  (25 \mu -131)+187}+187\right)}{9 (\mu -2) (\mu -1)}$ & $\frac{-5 \mu +\sqrt{\mu 
   (25 \mu -131)+187}+5}{9 (\mu -2)}$ \\\hline
$D$ & $\frac{9}{-5 \mu +\sqrt{\mu  (25 \mu -131)+187}+5}$ & $\frac{4 \left(\mu  \left(-25 \mu +\sqrt{\mu  (25 \mu
   -131)+187}+131\right)-7 \sqrt{\mu  (25 \mu -131)+187}-187\right)}{9 (\mu -2) (\mu -1)}$ & $\frac{5 \mu +\sqrt{\mu 
   (25 \mu -131)+187}-5}{9 (\mu -2)}$ \\\hline
$E$ & $\frac{-7 \mu +\sqrt{\mu  (49 \mu -242)+337}+7}{8 (\mu -2)}$ & $\frac{-7 \mu +\sqrt{\mu  (49 \mu -242)+337}+7}{12 (\mu
   -2)}$ &$ -\frac{\mu  \left(49 \mu +\sqrt{\mu  (49 \mu -242)+337}-242\right)-9 \sqrt{\mu  (49 \mu -242)+337}+337}{4
   (\mu -2) (\mu -1)}$ \\\hline
$F$ & $\frac{\mu  \left(-49 \mu +\sqrt{\mu  (49 \mu -242)+337}+242\right)-9 \sqrt{\mu  (49 \mu -242)+337}-337}{4 (\mu -2)
   (\mu -1)}$ &$ -\frac{7 \mu +\sqrt{\mu  (49 \mu -242)+337}-7}{8 (\mu -2)}$ &$ -\frac{7 \mu +\sqrt{\mu  (49 \mu
   -242)+337}-7}{12 (\mu -2)}$ \\\hline
$G$ & $ \frac{12}{\mu -1}-6$ &$ 2$ & $3$ \\\hline
$H $& $\frac{2}{\mu -2}$ &$ \frac{3}{\mu -2}$ &$ \frac{60}{\mu -1}-\frac{18}{\mu -2}-12 $\\\hline
$I$ & $\frac{2}{\mu -2} $& $\frac{3}{\mu -2}$ & $\frac{60}{\mu -1}-\frac{18}{\mu -2}-12 $\\\hline
    \end{tabular}}
    \caption{Eigenvalues of the Jacobian matrix evaluated at the equilibrium points of the system  \eqref{syst-1}, \eqref{syst-2} and  \eqref{syst-3}.}
    \label{tab:eigenvalues}
\end{table*}
\begin{figure*}
    \centering
    \includegraphics[scale=1.0]{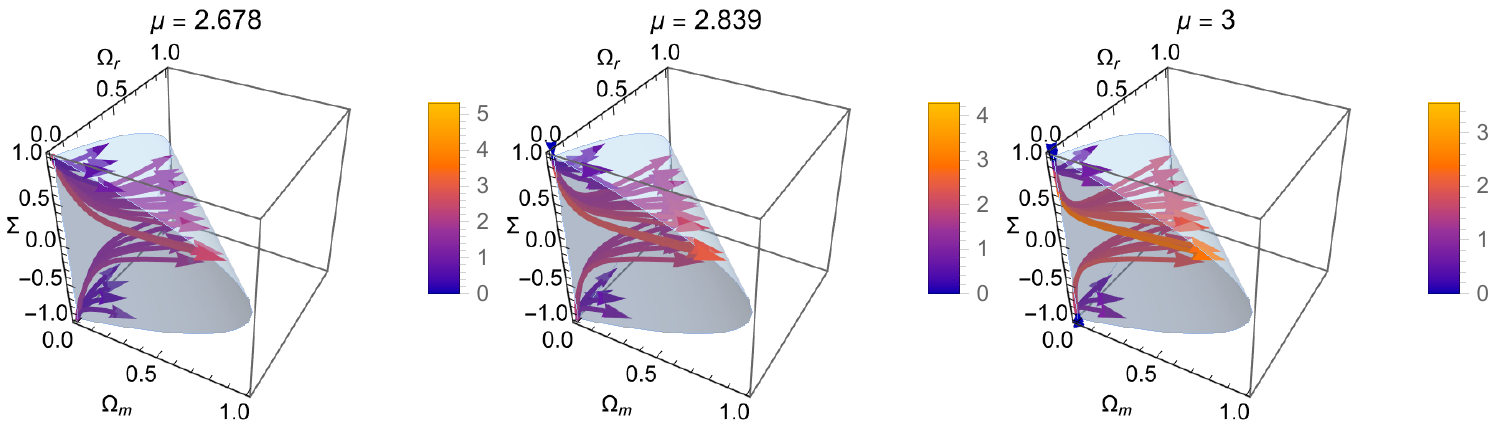}
    \caption{Phase space of the system \eqref{syst-1}, \eqref{syst-2} and  \eqref{syst-3} for some values of $\mu$. The gray surface  correspond to the boundary $1= \Sigma^2 + \sum_i\Omega_i $ corresponding to $A\rightarrow \infty$.}
    \label{fig:4}
\end{figure*}
To compare with the standard model we impose that the universe components are two ($n=2$), CDM  ($\rho_{1}=\rho_{\text{m}}, w_1=w_{\text{m}}=0$) and radiation  ($\rho_{2}=\rho_{\text{r}}, w_2=w_{\text{r}}=1/3$), which in our modified scenario, the dimensionless energy densities   evolve according to
\begin{small}
\begin{align}
    {\Omega _{\text{m}}^{\prime}} & =   {\Omega_{\text{m}}} \left[4-\Omega_{\text{m}}-\Omega_{\text{r}}-\Sigma ^2 \right. \nonumber \\
      & \left. -\frac{2 (\mu -2) \left(\Omega_{\text{m}}+\Omega_{\text{r}}+\Sigma ^2-1\right){}^2}{\mu -1}   -\frac{4 (\mu -4)
   \left(\Omega_{\text{m}}+\Omega_{\text{r}}+\Sigma ^2-1\right)}{\mu -1}\right], \label{syst-1}
\\
     {\Omega_{\text{r}}^{\prime}} & = {\Omega_{\text{r}}} \left[2-\frac{4}{3} \left(\Omega_{\text{m}}+\Omega_{\text{r}}+\Sigma
   ^2-1\right) \right. \nonumber \\
      & \left.  -\frac{2 (\mu -2) \left(\Omega_{\text{m}}+\Omega_{\text{r}}+\Sigma ^2-1\right){}^2}{\mu -1}    -\frac{4 (\mu -4)
   \left(\Omega_{\text{m}}+\Omega_{\text{r}}+\Sigma ^2-1\right)}{\mu -1}\right],  \label{syst-2}
\\
  \Sigma^{\prime} & = \frac{\Sigma  \left(1-\Omega_{\text{m}}-\Omega_{\text{r}}-\Sigma ^2\right)}{\mu -1}  \left[(\mu -2) \left( \Sigma ^2 +  \Omega_{\text{m}}+  \Omega_{\text{r}}\right)+2 \mu-7\right].  \label{syst-3}
\end{align}
\end{small}
The equilibrium points of the system  \eqref{syst-1}, \eqref{syst-2} and  \eqref{syst-3} are presented in Table \ref{tab:my_label}. 

In figure \ref{fig:4} is presented a phase space of the system \eqref{syst-1}, \eqref{syst-2} and  \eqref{syst-3} for $\mu \in\{2.678, 2.839, 3\}$. The possible late-time attractors are the equilibrium point $B$ with $\Omega _{\text{m}}=0$,  $\Omega _{\text{r}}=0$, $\Sigma=0$, $A=\mu -1$ and  $q=\frac{3 \mu -8}{\mu -1}$ which is a sink for $1<\mu <8/3$; the equilibrium point  $E$ with  $\Omega _{\text{m}}=-\frac{\mu +\sqrt{\mu  (49 \mu -242)+337}-9}{4 (\mu -2)}$, $\Omega _{\text{r}}=0$, $\Sigma=0$, $A= \left(-5 \mu +\sqrt{\mu  (49 \mu
   -242)+337}+17\right)/6$, $q=-\frac{\sqrt{49 \mu ^2-242 \mu +337}+\mu -9}{8 (\mu -2)}$. It is a sink for $\mu >8/3$; and 
the equilibrium point $F$ with  $\Omega _{\text{m}}=\frac{-\mu +\sqrt{\mu  (49 \mu -242)+337}+9}{4 (\mu -2)}$, $\Omega _{\text{r}}=0$, $\Sigma= 0$, $A= \left(-5 \mu -\sqrt{\mu  (49 \mu
   -242)+337}+17\right)/6$ and $q=\frac{\sqrt{49 \mu ^2-242 \mu +337}-\mu +9}{8 (\mu -2)}$, which is a sink for $\mu >2$.

\begin{figure}
    \centering
    \includegraphics[width=0.45\textwidth]{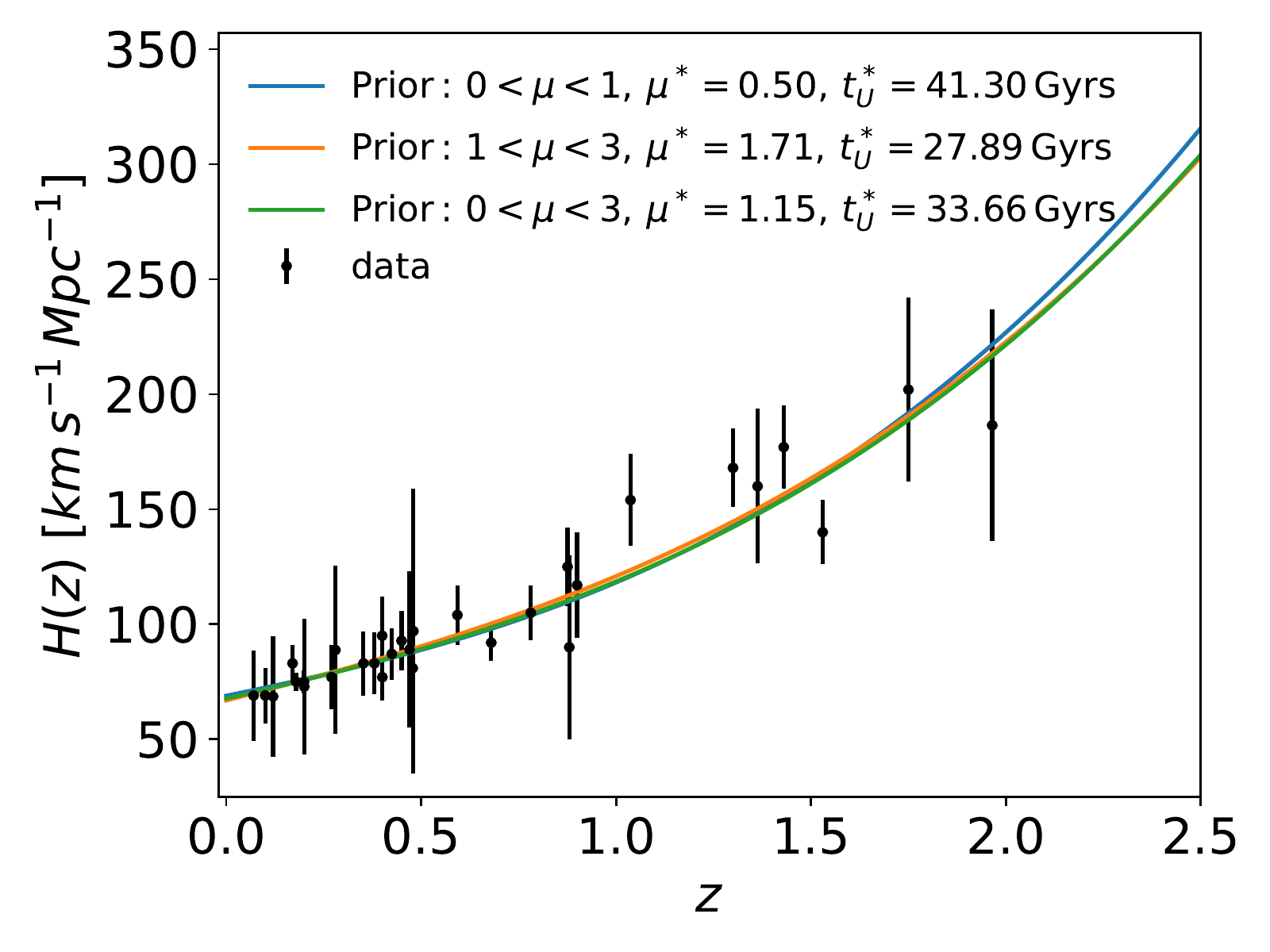}
    \caption{$H(z)$ reconstruction using Eqs. \eqref{eq:c1_newH}, \eqref{tz} and \eqref{alternativeE} and best-fit values ($\mu^*$, $t_U^*$) for different priors on $\mu$.}
    \label{fig:NewH}
\end{figure}

To close the ideas presented in this paper, we finally show alternative calculations (see Subsec. \ref{FracFriedmann} and Appendix \ref{AlternativeApp}) in order to make more efficient the numerical computation, in particular with the free parameter constraints. In this case we invert \eqref{tz} and substitute into 
\eqref{alternativeE} and
\eqref{Alternative-q}, in order to obtain an alternative form for $H(z)$ and $q(z)$ where now is used a new notation for the free parameters $h$, $\mu^*$ and $t_U^*$. In this formulation, we have, for the different priors, the best-fit parameters summarized in Fig. \ref{fig:NewH}. This new approach works for FLRW and Bianchi I models and deserves further investigation.

\section{Summary and Discussions} \label{SD}

We study the recent proposition of fractional cosmology to elucidate if the theory is capable of reproducing the observed dynamics of the Universe, in specific, if it is capable of predicting the Universe's acceleration and giving some clues about the fundamental nature of the dark energy. We implement constraints through cosmic chronometers, Type Ia Supernovae, and joint analysis and summarized our results in Fig. \ref{fig:contours} and Table \ref{tab:bestfits}. The fractional parameter prefers $\mu=2.839^{+0.117}_{-0.193}$ for a joint analysis which suggests a solid presence of fractional calculus in the dynamical equations of cosmology; however, it generates crucial differences as it is possible to observe from Figs \ref{fig:Hz_and_qz}. Of course, we expect this behaviour to have an accelerated Universe at late times. From one side, the term $(1-\mu)H/t$ acts like an extra source of mass, closing the Universe and not allowing the observed dynamics, in particular, the Universe acceleration at late times if $\mu<2$, but, for $\mu>2$, we can have an accelerated power-law solution. Furthermore, from Figs. \ref{fig:Hz_and_qz} it is possible to notice that the fractional constant $f$ can act like the object that causes the Universe acceleration. It is possible to observe from $H(z)$ and $q(z)$ essential differences when we compare them with the standard model, mainly at high redshifts. In addition, the jerk parameter also shows that the causative of the Universe acceleration is not a cosmological constant because, at $z=0$, the fractional parameter does not converge to $j=1$; this coincides with recent studies that suggest that it is not a $\Lambda$ the cause of the Universe acceleration \citep{Zhao:2017cud}.
Moreover, the Universe's age obtained under this scenario is $t_U=33.617^{+3.411}_{-4.511}$ Gyrs based on our Joint analysis, around $2.4$ times larger than the age of the Universe expected under the standard paradigm. However, this value does not contradict the minimum bound expected for the universe age imposed by globular clusters, and, as far as we know, the maximum bound does not exist and is model-dependent. Finally, we observe a trend of $H_{0}$ to the value obtained by SH0ES \citep{Riess:2019cxk} at current times, and in agreement with Planck's value \citep{Planck:2018} for $z\lesssim1.5$. However, a discrepancy between both values in the region $1.5<z<2.5$ holds, such that $H_0$ tension is not fully resolved.
\newline
Additionally, we have presented a dynamical system and stability analysis to explore the phase-space under the assumption of different $\mu$ parameters. This formulation enabled us to achieve a good understanding of the global structure of the reduced phase space. One late-time attractor have $a(t)= \left(t/t_U\right)^{\frac{1+\mu}{3}}$, which is physical for  $1\leq \mu <\frac{5}{2}$. Evaluating the dimensionless energy densities $\Omega_{\text{m}}$, $\Omega_{\text{r}}$ of matter and radiation, and the Hubble parameter, we identify $P_3$ with a power-law (decelerated) late-time attractor for $\mu<2$. 
Moreover, for $\mu>1$ exists an additional point not present in GR with  $a(t)= \left(t/t_U\right)^{\mu-1}$ which can be  a source for $\mu <\frac{7}{4}$ or a saddle for $\frac{7}{4}<\mu <2$, or a sink for $\mu >2$. Evaluating the dimensionless densities $\Omega_{\text{m}}$, $\Omega_{\text{r}}$ of matter and radiation, and the Hubble parameter, we identify $P_4$ with an accelerated power-law if $\mu<1$ or $\mu>2$,  or decelerated power-law   if $1<\mu <2$. 
\newline 
For the Bianchi I metric, the possible late-time attractors are all isotropic ($\Sigma=0$). They are the equilibrium point $B$ which is a sink for $1<\mu <8/3$; the equilibrium point  $E$, which  is a sink for $\mu >8/3$; and 
the equilibrium point $F$, which is a sink for $\mu >2$.
   
 Moreover,  the new approach of fractional calculus opens new windows to affront calculations that traditional calculus can not resolve. For example, the problem of determining the energy density value for the cosmological constant can be attached to this approach or even applied to the standard model field to make its calculations efficient. As we demonstrate in this research, fractional calculus contributes with a constant that acts as the causative of the Universe's acceleration without the need to add no natural term into the field equations. Indeed, if we had written the Einstein Field equations in the fractional setup, the Friedmann equations naturally contained a constant term, predicting the existence of a late time Universe in acceleration contrary to the standard approach. Moreover, as it is possible to observe, the model presents differences in comparison with the standard model at high redshifts being the smoking gun to differentiate among the theories.

Finally, from a mathematical perspective, it is expected that many physical phenomena and systems are better described by fractional differential equations rather than the equivalent integer-order equations. Since the set of integers is a set of measure zero, and the real numbers are a set of measure one, nature prefers non-integer dimensions and parameters. Therefore, we recommend that the community study other approaches like this presented in the paper to understand the Universe's acceleration with another mathematical background. For example, in fractional calculus, the mathematical richness generates the  $\Lambda$-like term originated in the corrections due to the fractional index $\mu$ of the fractional derivative, which could resolve the energy density problem. In future studies, it is possible to aboard the fractional Einstein equation using linear perturbations theory to understand acoustic peaks of the CMB, the power spectrum and inflation. However, this will be presented elsewhere.

\section*{Acknowledgments}
We thank the anonymous referee for thoughtful remarks and suggestions. M.A.G.-A. acknowledges support from c\'atedra Marcos Moshinsky and Universidad Iberoamericana for support with the S.N.I. grant; G.F.A. acknowledges support from DINVP and Universidad Iberoamericana. A.H.A. thanks to the support from Luis Aguilar, Alejandro de Le\'on, Carlos Flores, and Jair Garc\'ia of the Laboratorio 
Nacional de Visualizaci\'on Cient\'ifica Avanzada. G.L. was funded by  Vicerrectoría de Investigación y Desarrollo Tecnológico (Vridt) at U.C.N. and through Concurso De Pasantías De Investigación Año 2022, Resolución Vridt N 040/2022 under the Project ``The Hubble constant tension: some ways to
alleviate it'' and through Resolución Vridt No. 054/2022. G.L. acknowledges the invitation of organizers of the conference ``Tensions in Cosmology'', held on Sep 7 - Sep 12, 2022 - in Corfu, Greece, where part of these results was presented. J.M. acknowledges the support from  ANID REDES 190147.

\section*{Data Availability}
The data underlying this article were cited in Section \ref{sec:constraints}.

\bibliographystyle{mnras}
\bibliography{main}

\begin{strip}
\begin{appendix}

\section{Deceleration parameter as a closed formula of redshift.}
Here we compute the deceleration parameter, using Eq. \eqref{q(z)}, which yields to 
\begingroup\makeatletter\def\f@size{9}\check@mathfonts
\begin{align}
    &q(z)=-1 + \frac{(z+1)}{E(z)} \Bigg\{\frac{2 f^2 F(z)^2}{(\mu -1) (z+1) E(z)} \nonumber \\
    & +\frac{(E(z)+f F(z)) \Big(-\frac{12 f^3
   F(z)^{\frac{4 \mu }{3}+\frac{11}{3}}}{(\mu -1) (z+1) E(z)}+\frac{6 f \Omega_{0\text{m}} (z+1)^2
   F(z)^{\frac{\mu +8}{3}}}{E(z)}+\frac{8 f \Omega_{0\text{r}} (z+1)^3 F(z)^3}{E(z)}+9
   \Omega_{0\text{m}} (z+1)^2 F(z)^{\frac{\mu +5}{3}}+12 \Omega_{0\text{r}} (z+1)^3 F(z)^2\Big)}{6 \left(f^2
   F(z)^{\frac{4 (\mu +2)}{3}}+\Omega_{0\text{m}} (z+1)^3 F(z)^{\frac{\mu +5}{3}}+\Omega_{0\text{r}} (z+1)^4
   F(z)^2\right)}\Bigg\}, \label{DecelerationFinal}
\end{align}
\endgroup
where $E(z)$ function is given by \eqref{FriedmannFinal1}.
\section{Alternative $E$ and deceleration parameter} \label{AlternativeApp}
Alternative expressions for $E$ and the deceleration parameter that makes use of the exact solution of \eqref{NewIntegrableH} given by \eqref{solution} are
\begin{small}
\begin{align}
    E(t) & =  \frac{1}{H_0} \left[\frac{9-2 \mu }{6 t}+\frac{\sqrt{8 \mu  (2 \mu -9)+105} \left(1-\frac{2 c_1}{t^{\sqrt{8 \mu  (2 \mu
   -9)+105}}+c_1}\right)}{6 t} \right], \label{alternativeE}
\end{align}    
\end{small}
and 
\begin{small}
\begin{align}
  q(t)=  -1+\frac{-6 c_1{}^2 \left(2 \mu +\sqrt{8 \mu  (2 \mu -9)+105}-9\right)+6 \left(-2 \mu +\sqrt{8 \mu  (2 \mu -9)+105}+9\right) t^{2 \sqrt{8 \mu  (2 \mu
   -9)+105}}-24 c_1 (\mu  (8 \mu -35)+48) t^{\sqrt{8 \mu  (2 \mu -9)+105}}}{\left(\left(-2 \mu +\sqrt{8 \mu  (2 \mu -9)+105}+9\right) t^{\sqrt{8 \mu  (2
   \mu -9)+105}}-c_1 \left(2 \mu +\sqrt{8 \mu  (2 \mu -9)+105}-9\right)\right){}^2}. \label{Alternative-q}
\end{align}
\end{small}
where $c_1$ is defined by \eqref{H(t)}, and the relation between $t$ and $z$ is obtained by inverting \eqref{tz}.
\section{Unstable manifold of $P_2$}
\label{app2}
To find the unstable manifold of $P_2$, we define
\begin{small}
\begin{align}
   u & =-\frac{(2-5 \mu )^2 x_1}{64 \mu ^2-68 \mu -104}, \quad
   v  =  -\frac{\mu }{3}+A+\frac{21 (5
   \mu +34) x_1}{64 (\mu  (16 \mu -17)-26)}+\frac{25 x_1}{64}-\frac{1}{6}.\label{B2}
\end{align}
\end{small}
The graph
\begin{equation}
    \left\{(u,v): v=g(u), g(0)=g'(0)=0\right\}  \label{B3}
\end{equation}
 locally gives the unstable manifold of $P_2$, where $g$ satisfies the differential equation
\begin{small}
\begin{align}
& 4 ((17-16 \mu ) \mu +26) (2-5 \mu )^2 u (-\mu +6 g(u)+6 u+4) (5 \mu +6 g(u)+6 u-2) \nonumber \\
& \times \Bigg[3 (\mu+1)
   -\left(\frac{4 ((17-16 \mu ) \mu +26) u}{(2-5 \mu )^2}+25\right) \left(\frac{\mu
   }{3}+g(u)+u+\frac{1}{6}\right)+\frac{2 (2 \mu +6 g(u)+6 u+1) (5 \mu +15 g(u)+15 u-2)}{5 \mu +6 g(u)+6
   u-2} \nonumber \\
   & +\frac{6 \left(\frac{\mu }{3}+g(u)+u+\frac{1}{6}\right) \left(2 \left(\frac{4 ((17-16 \mu ) \mu +26)
   u}{(2-5 \mu )^2}+1\right) \left(\frac{\mu }{3}+g(u)+u+\frac{1}{6}\right)-3\right)}{\mu -6 g(u)-6
   u-4}-\frac{4 ((17-16 \mu ) \mu +26) \mu  u}{(2-5 \mu )^2}+\frac{4 ((17-16 \mu ) \mu +26) u}{(2-5
   \mu )^2}\Bigg] g'(u) \nonumber \\
   & +\left(64 \mu ^2-68 \mu -104\right) \Bigg\{-36 u^4 (5 \mu  (2 \mu  (56 \mu
   -45)-201)+60 (\mu  (16 \mu -17)-26) g(u)+166) \nonumber \\
   & -6 u^3 ((5 \mu -2) (\mu  (\mu  (400 \mu
   -459)-138)-532)+6 g(u) (\mu  (4 \mu  (340 \mu -207)-2649)+72 (\mu  (16 \mu -17)-26) g(u)+470)) \nonumber\\
   & +u^2
   \Big[12 g(u) (3 g(u) (\mu  (52 (3-20 \mu ) \mu +2403)+36 ((17-16 \mu ) \mu +26) g(u)-466)-(5
   \mu -2) (\mu  (\mu  (320 \mu -393)+132)-464)) \nonumber \\
   & -(2-5 \mu )^2 (\mu  (4 \mu  (20 \mu
   -9)-69)-218)\Big] \nonumber \\
   & +u g(u) \Big[18 g(u) \left(-((5 \mu -2) (\mu  (\mu  (80 \mu -79)+182)-156))-6 g(u)
   \left(\mu  (4 \mu  (20 \mu +31)-293)+\left(32 \mu ^2-34 \mu -52\right) g(u)+62\right)\right) \nonumber \\
   & -(2-5
   \mu )^2 (\mu  (80 (\mu -3) \mu +357)-116)\Big] \nonumber \\
   & +(2-5 \mu )^2 g(u) (4 \mu -6 g(u)-7) (2 \mu
   +6 g(u)+1) (5 \mu +6 g(u)-2) -648 (\mu  (16 \mu -17)-26) u^5\Bigg\}=0.\label{ODE}
\end{align}
\end{small}
\end{appendix}
\end{strip}

\bsp	
\label{lastpage}

\end{document}